\documentclass[
 reprint,
superscriptaddress,
 amsmath,amssymb,
 aps,
prb,
citeautoscript
]{revtex4-1}

\usepackage{amssymb}
\usepackage{graphicx}
\usepackage{xspace}
\usepackage{dsfont}

\usepackage{graphicx}
\usepackage{dcolumn}
\usepackage{bm}
\usepackage[svgnames]{xcolor}
\usepackage[colorlinks=True,linkcolor=DarkRed,citecolor=Blue,urlcolor=MediumBlue,pdfstartview=FitH,bookmarks=False,pdfpagemode=UseNone]{hyperref}

\usepackage{xspace}
\newcommand{\indexladder}{\mathcal{L}}
\newcommand{\indexdimer}{\mathcal{D}}
\newcommand{\indexchain}{\mathcal{C}}
\newcommand{\namemagn}{\textit{studied magnetic point}\xspace}
\newcommand{\TLL}{TLL\xspace}
\newcommand{\TDMRG}{T-DMRG\xspace}
\newcommand{\DMRG}{DMRG\xspace}
\newcommand{\XXZ}{XXZ\xspace}

\newcommand{\CHAINAX}{0.135}
\newcommand{\CHAINBX}{0.021}
\newcommand{\CHAINAZ}{0.09}
\newcommand{\CHAINK}{0.745}
\newcommand{\DIMERAX}{0.1469}
\newcommand{\DIMERBX}{0.0135}
\newcommand{\DIMERAZ}{0.082}
\newcommand{\DIMERK}{0.754}
\newcommand{\LADDERAX}{0.166}
\newcommand{\LADDERBX}{0.007}
\newcommand{\LADDERAZ}{0.078}
\newcommand{\LADDERK}{0.85}

\newcommand{\average}[1]{\left\langle #1 \right\rangle}
\newcommand{\conj}[1]{\overline{#1}}
\newcommand{\norm}[1]{\left\lVert#1\right\rVert}
\newcommand{\ket}[1]{\left|#1\right\rangle}
\newcommand{\bra}[1]{\left\langle#1\right|}
\newcommand{\abs}[1]{\left|#1\right|}

\renewcommand{\Im}{\operatorname{Im}}

\usepackage{ stmaryrd }
\newcommand{\pseudoS}{\mathds{S}}
\newcommand{\pseudoUp}{\mathbin{\rotatebox[origin=c]{90}{$\rightarrowtriangle$}}}
\newcommand{\pseudoDown}{\mathbin{\rotatebox[origin=c]{-90}{$\rightarrowtriangle$}}}

\newcommand{\DIMPY}{DIMPY\xspace}
\newcommand{\BPCC}{BPCC\xspace}
\newcommand{\BPCB}{BPCB\xspace}
\newcommand{\BPCCcool}{bis-piperidinium copper tetrachloride\xspace}

\begin{document}

\title{Low-dimensional correlations under thermal fluctuations}

\author{N. Kestin}
\affiliation{Department of Quantum Matter Physics, University of Geneva,
1211 Geneva, Switzerland}

\author{T. Giamarchi}
\affiliation{Department of Quantum Matter Physics, University of Geneva,
1211 Geneva, Switzerland}

\date{\today}

\begin{abstract}
  We study the correlation functions of quantum spin $1/2$ two-leg ladders at
  finite temperature, under a magnetic field, in the gapless phase at
  various relevant temperatures $T\neq 0$, momenta $q$, and
  frequencies $\omega$. We compute those quantities using the time-dependent density-matrix renormalization group (T-DMRG) in an
  optimal numerical scheme. We compare these correlations with the
  ones of dimerized quantum spin chains and simple spin chains, that
  we compute by a similar technique.  We analyze the intermediate
  energy modes and show that the effect of temperature leads to the
  formation of an essentially dispersive mode corresponding to the
  propagation of a triplet mode in an incoherent background, with a
  dispersion quite different from the one occurring at very low
  temperatures. We compare the low-energy part of the spectrum with
  the predictions of the Tomonaga-Luttinger liquid field theory at
  finite temperature. We show that the field theory describes in a
  remarkably robust way the low-energy correlations for frequencies or
  temperatures up to the natural cutoff (the effective dispersion) of
  the system. We discuss how our results could be tested in,
  e.g., neutron-scattering experiments.
\end{abstract}


\maketitle

\section{Introduction}\label{sec:introduction}

The study of a strongly correlated system is of crucial importance for
both the cold atom and the condensed matter communities. In
particular, both are able to provide experimental realizations with
well-controlled microscopic Hamiltonians using either optical
lattices\cite{bloch_MB_cold,cazalilla_condensedtocold_giam} or quantum
magnets\cite{assa_magnetism_book}.  On the theory side, going from the
knowledge of the microscopic Hamiltonian to the calculation of the
correlations, which can be compared to experimental measurements, is
of course a considerable challenge.

One class of systems which presents a very rich set of phases,
depending on the precise microscopic interactions, is the one of
quantum one-dimensional or quasi-one-dimensional magnets
\cite{giam_1d_book}. Indeed, such systems possess ground states ranging
from quasi-long-range magnetic order to spin liquids. The coupling of
several one-dimensional chains as ladders leads to a very rich phase
diagram as a function of the number of legs \cite{rice_quasi1vs2}.
The correlations in these systems can be probed by, e.g., inelastic
neutron-scattering (INS)\cite{ins_epfl_psi_book} or nuclear magnetic
resonance (NMR)\cite{berthier_nuclear_2017} experiments, giving a very
complete access to the spatial or time dependence of the spin-spin
correlations.

In such systems the precise knowledge of the microscopic Hamiltonian
allows thus for a drastic test of the theoretical methods used to
compute the correlations.  However, computing the correlation
analytically, by methods such as Bethe-Ansatz \cite{caux_XXZ} has only
proven possible at zero temperature. Comparison with experiments could
thus be done for probes, such as neutrons, when the energy of the
probe is much larger than the temperature
\cite{thielemann_spectrum_spinladder, thielemann_bpcp_neutron_ladder}.
Numerical methods, such as the density-matrix renormalization group
(DMRG)
\cite{white_dmrg_article1_breakthrough,white_dmrg_article2_breakthrough,
  vidal_tebd_precursor_pure_state_dynamics_breakthrough,daley_kollath_schollwock_tdmrg_pure_state_dynamics_breakthrough,
  vidal_tebd_pure_state_dynamics_breakthrough,white_feiguin_tdmrg_pure_state_breakthrough,schollwock_dmrg_long_notes},
allowed for a direct calculation of the zero-temperature correlations
that could be successfully compared with experiments for ladder
systems \cite{bouillot_long_ladder,schmidiger_neutrons_bound_spinons}.

An important challenge is of course to properly incorporate the finite
temperature effects. For temperatures much lower than the magnetic
exchanges in the problem this can be accomplished by using a
combination of the field theory description, such as the
Tomonaga-Luttinger liquid (\TLL) theory \cite{giam_1d_book}, and
numerics to get an essentially quantitative finite temperature
description, which could be successfully compared to experiments
\cite{klanjsek_bpcp, bouillot_long_ladder,
  schmidiger_dimpy_neutrons}. However, this description breaks down
when the temperature becomes comparable to the exchanges or close to
a quantum critical point
\cite{sachdev_qcp_book,sachdev_qcp_antiferro_finiteT}, and it is
desirable to have a direct way to quantitatively compute the
correlations at finite temperature.

Fortunately such a method is provided by the DMRG, which can be used
to compute the finite temperature dynamical correlations at the
expense of much more heavy
calculations\cite{verstraete_garciaripoll_cirac_tdmrg_dissipative_breakthrough,
  vidal_zwolak_tebd_temperature,
  feiguin_white_Tdmrg_just_after,barthel_tdmrg_optimized_schemes}. This
program has been carried out with success for spin-$1/2$
chains\cite{barthel_xxz_Tdmrg} where it allowed one in particular to
analyze the surroundings of the quantum critical point close to
saturation and neutron experiments\cite{noam_quasi1D}.  Spin-1
single-ion anisotropy\cite{honecker_spinone_Tdmrg,lange_spinone_tdmrg}
and dimerized chains\cite{klyushina_dimer_Tdmrg} could be analyzed at
finite temperature. For the dimers both NMR
\cite{coira_giam_kollath_spin_lattice_relax_nmr_low_t} and the neutron
scattering\cite{coira_temperature_dimer} could be computed, allowing one to
investigate the broadening effects due to the temperature on the
spectrum\cite{damle_temperature_spin_dynamics}.

We investigate in the present paper the thermal effects on the
spin-spin correlations of a two-leg ladder system.  On the theory side
this allows for a comparison between the two-leg ladder and the
dimerized systems. On
the more experimental side this is stimulated by recent INS experiments
done in weakly coupled spin-$1/2$ ladders which were done close to a
quantum critical point \cite{blosser_qcp_ladder} or the existence of
compounds with
relatively small magnetic exchange such as \BPCCcool
(\BPCC)\cite{ward_bpcc_compound}, for which we can expect the effects
of temperature to be \textit{a priori} more important.  From the technical point
of view the two-leg ladders are more challenging due to the greater
entanglement compared to either spin chains or dimers. In this paper
we will mostly focus on the comparison of the thermal effects between
the ladders, dimers, and chains. We also compare the direct numerical
calculations with the field theory description at finite temperature
in order to have a feeling of the range of validity of the field
theory description, in a spirit similar to what was done previously
for NMR \cite{coira_giam_kollath_spin_lattice_relax_nmr_low_t}.

The plan of the paper is as follows.  In Sec.~\ref{sec:models}, we
introduce the low-dimensional models that we will study in the paper. We then explain details about the numerical algorithm for
the measure of the low-dimensional correlations in
Sec.~\ref{sec:method}. We then present in Sec.~\ref{sec:channels} the dynamical
structure factors  of the various models at
different temperatures that INS experiments can
measure also. We finally compare the low-energy spectrum with some
analytical field theory in Sec.~\ref{sec:analytic} and discuss the
deviation from bosonization expectations.

\section{\label{sec:models}Models}
In this paper, we focus on three classes of problems made of coupled
spin-$1/2$, namely, (i) ladder systems $\indexladder$ made of two
coupled spin chains, (ii) weakly dimerized chains $\indexdimer$, and (iii)
$\Delta=\frac{1}{2}$ anisotropic \XXZ chains $\indexchain$.

\subsubsection{Ladder $\indexladder$}\label{sec:ladders}
\begin{figure}[th]
  \centering
   \includegraphics[scale=0.3]{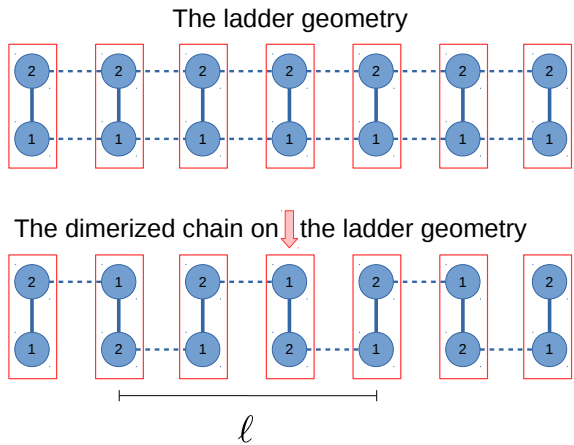}
   \caption{\label{fig:models} Weakly coupled dimer and two-leg ladder
     representation. The index $\eta$ corresponds for the ladder to
     the bottom or upper leg. For the dimer, $\eta$ corresponds to the
     left or right strong bond cell, thus the labels shuffle when
     mapped on the two-leg ladder geometry. We add an arrow on the middle cell
     to visualize the symmetry when we inverse the dimerized chain.}
\end{figure}

We consider a two-leg ladder system with spins coupled by
antiferromagnetic Heisenberg couplings on rungs and legs (see
Fig.~\ref{fig:models})
\begin{equation}
  \label{eq:hamiltonian_ladder}
  H_\indexladder = J_\parallel   \sum_{\ell,\eta} \mathbf{S}_{\ell,\eta} \cdot \mathbf{S}_{\ell+1,\eta} + J_\perp \sum_{\ell} \mathbf{S}_{\ell,1} \cdot \mathbf{S}_{\ell,2} - h^z \sum_{\ell,\eta} S^z_{\ell,\eta}
\end{equation}
with stronger rung coupling where $\mathbf{S}_{\ell,\eta}$ denotes a
spin-$1/2$ at rung $\ell$ on leg $\eta\in \{1,2\}$. The spin-$1/2$
$\mathbf{S}=(S^x,S^y,S^z)=\frac{\hbar}{2}(\sigma^1,\sigma^2,\sigma^3)$
can be decomposed in lowering and raising operators
$S^\pm=S^x\pm i S^y$, where we denote by $\sigma^i$ the Pauli
matrices, $i \in \{ 1,2,3\}$. Note that the coupling values are given
in \eqref{eq:coupling_ladder}.

\subsubsection{Dimer $\indexdimer$}\label{sec:dimers}
If we remove alternatively the weak bonds along the ladder (see
Fig.~\ref{fig:models}) and map the model to a chain we get a dimerized
chain of alternative bonds. For an even number of sites $N$, we always
have $\frac{N}{2}$ strong bonds $J_s$ and $\frac{N}{2}-1$ weak bonds
$J_w$.  The model is thus
\begin{equation}
  \label{eq:hamiltonian_dimer}
  H_\indexdimer = \sum_{n=1}^{N-1} (J - (-1)^{n} \delta J) \mathbf{S}_n \cdot \mathbf{S}_{n+1} -h^z \sum_{n=1}^{N} S^z_n
\end{equation}
starting with a strong bond at each border $J_s=J+\delta J$ and
alternating with the weak bonds $J_w=J-\delta J$ along the chain. The
coupling values are given in \eqref{eq:coupling_dimer}.

\subsubsection{Spin-Chain mapping - $\Delta=\frac{1}{2}$ \XXZ chain $\indexchain$}\label{sec:spinchainmapping}
Both previous models can be mapped to an anisotropic single spin chain
in some regime of parameters when studying the low-energy
behavior\cite{bouillot_long_ladder}. If the magnetic field and
temperature are such that we neglect the triplet $\ket{t^0}$ and
$\ket{t^-}$ population, one can identify a spin-chain behavior in the
critical interplay between $\ket{t^+}$ and $\ket{s}$.

We introduce the pseudo-spin-$1/2$ $\vec{\pseudoS}$ in the basis
$\ket{\pseudoUp} \equiv \ket{t^+} $,
$\ket{\pseudoDown} \equiv \ket{s} $ mapped by
$S^\pm_{\eta} \equiv \frac{\eta}{\sqrt{2}} \pseudoS^\pm$ and
$S^z_{\eta} \equiv \frac{1}{4} \left( 1+ 2\pseudoS^z \right)$ in the
singlet-triplet crossing region.  The mapping leads to a spin-$1/2$
\XXZ chain with $\Delta=\frac{1}{2}$ anisotropy
\begin{equation}
  \label{eq:hamiltonian_chain}
  H_\indexchain  = J \sum_{\ell=1}^{N-1} \left( \pseudoS^x_\ell \pseudoS^x_{\ell+1} + \pseudoS^y_\ell \pseudoS^y_{\ell+1} + \frac{1}{2}  \pseudoS^z_\ell \pseudoS^z_{\ell+1} \right)   - h^z_{eff}  \sum_{\ell=1}^{N} \pseudoS^z_\ell
\end{equation}
The spin-chain mapping fixes the following microscopic parameters for
the \XXZ model
\begin{enumerate}
\item ladder :  $J \equiv J_\parallel$ and
$h^z_{eff} = h^z - J_\perp -\frac{J_\parallel}{2}$,
\item dimer : $J \equiv \frac{-J_w}{2}$ and $h^z_{eff} = h^z - J_s -\frac{J_w}{4}$.
\end{enumerate}
Although these models can be studied independently we consider them
here in the regime where their low-energy properties are roughly
equivalent. We consider spin chains close to zero magnetization
$m_{\indexchain}=0$, which means that both the two-leg ladder and the dimer
are at a magnetization around half saturation $m_{\indexladder}=0.5$
and $m_{\indexdimer}=0.25$ at $T=0$. We call this point in the paper
the {\textbf{\namemagn}} for simplicity.

For the numerical study we fix the ratio of coupling constants of the
ladder to values corresponding roughly to the compound
\BPCC\cite{ward_bpcc_compound,ward_bpcc_singlet,
  ryll_gruneisen_bpcc,tajiri_magnetisation_bpcc} namely
\begin{equation}
  \label{eq:coupling_ladder}
  \begin{split}
    J_\parallel &= 0.39~J_\perp  \\
    J_\perp &= 1
  \end{split}
\end{equation}
In the same way for the dimer system we have
\begin{equation}
  \label{eq:coupling_dimer}
  \begin{split}
    J_w &= J-\delta J = 0.39~J_s  \\
    J_s &= J+\delta J = 1
  \end{split}
\end{equation}
The corresponding values for the magnetic field are respectively
$h^z \simeq 1.28~J_\perp$ for the two-leg ladder and
$h^z \simeq 1.148~J_s$ for the dimer.  We will consider these values in the rest of the
paper.  We will not discuss the correction of the constant and
the boundary terms in this paper.

One can see how the spin-chain mapping manifests on both models
looking at Figs.~\ref{fig:channels_ladder025} and
\ref{fig:channels_ladder059} compared with
Figs.~\ref{fig:channels_dimer025} and \ref{fig:channels_dimer059}.

\section{\label{sec:method} Method \TDMRG ($T \neq 0$)}

We implement in this paper a time-dependent density-matrix
renormalization group (\TDMRG) procedure
\cite{verstraete_garciaripoll_cirac_tdmrg_dissipative_breakthrough,
  vidal_zwolak_tebd_temperature,feiguin_white_Tdmrg_just_after,barthel_tdmrg_optimized_schemes},
a method based on the earlier \DMRG algorithm
\cite{white_dmrg_article1_breakthrough,white_dmrg_article2_breakthrough,
  vidal_tebd_precursor_pure_state_dynamics_breakthrough,daley_kollath_schollwock_tdmrg_pure_state_dynamics_breakthrough,
  vidal_tebd_pure_state_dynamics_breakthrough,white_feiguin_tdmrg_pure_state_breakthrough,schollwock_dmrg_long_notes}.

\begin{figure}[h]
  \centering
  \includegraphics[scale=0.55]{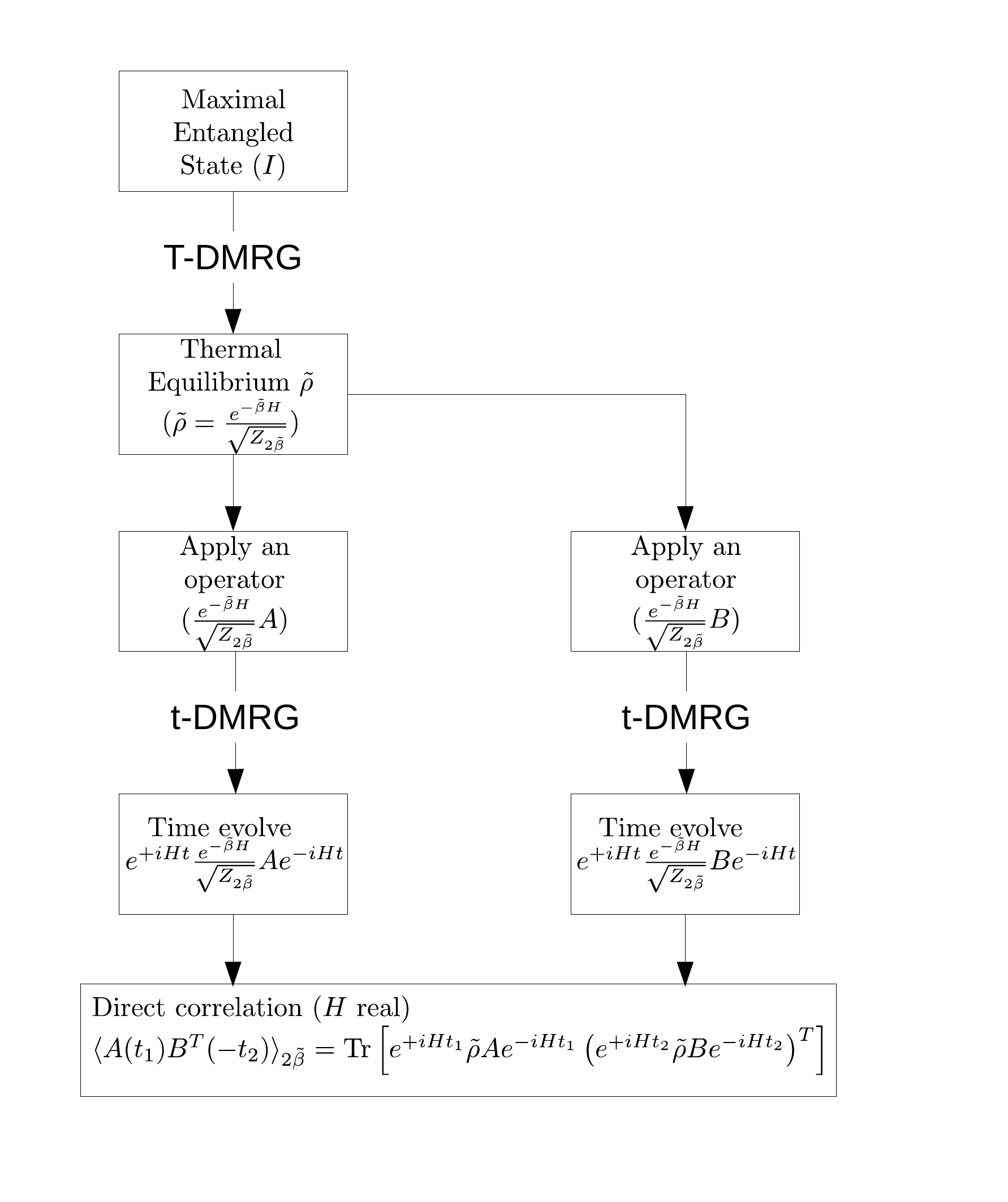}
  \caption{\label{fig:schemeC} Simulation and measurement of direct
    correlations by evolving two observables in time to $t_1$ and
    $t_2$ at finite temperature using the \TDMRG algorithm.  An
    optimal scheme, explained in
    Ref.~\onlinecite{barthel_tdmrg_optimized_schemes}, consists in
    evolving separately the two observables. This scheme requires
    storing all intermediate steps and contracting them at the end
    (see text). We use this scheme for the ladders since it can in
    principle double the resolution.  The standard scheme consists in
    evolving only one observable in time ($t_2=0$) and in the present
    paper we use it both for the dimers and for the chains. Please note,
    that in our notation $\beta = 2\tilde{\beta}$.}
\end{figure}

The method is schematically represented in Fig.~\ref{fig:schemeC}.
The time or imaginary time evolution follows the Suzuki-Trotter
decomposition\cite{suzuki_fractalization_exp_gates,
  mclachlan_leapfrog_exp}. In this paper we used a fourth-order
decomposition -- that expands the exponentials in terms of gates which
now can converge to the thermal equilibrium function
$\frac{1}{Z}e^{-\beta H}$.  One introduces hierarchical matrices for
tensors that increase the amount of information stored in the system
based on the local quantum
numbers\cite{singh_qn_dmrg_tensornetwork,hubig_qn_simple} and the
global conservation rules.

Both \DMRG and \TDMRG algorithm have the same complexity limit in term
of the bond dimension $\chi$ of the matrices -- during updates after
application of above-mentioned gates. The singular value
decomposition appears to scale\cite{vidal_tebd_pure_state_dynamics_breakthrough}
as $O(\chi) \sim A \chi^3$ but with different prefactors $A$, which
increases from $d^3$ for \DMRG to $d^6$ for \TDMRG, where $d$ is the
number of local degrees of freedom.  A similar scaling applies to the
memory case in which $O(\chi) \sim A \chi^2$ with $A \propto d^2$ for \DMRG
and $A \propto d^4$ for \TDMRG. For the case of ladders, due to the
effective longer range of the couplings, if one represents the system
as a chain, the gates have to be applied further which increases
further the complexity. This is related to the fact that \DMRG is most
efficient for quantum problems with a sufficiently low amount of
relevant information\cite{eisert_entanglement_arealaw} and thus
particularly for the low-dimensional problems.

\subsection{\TDMRG and a close to optimal scheme}\label{sec:schemeC}

Since the complexities of the two-leg ladder and of the dimer are different due
to the longer range of the coupling (see Fig.~\ref{fig:models}), one
needs to restrict the bond dimension $\chi$ according to the
problem. We use in this paper values of $\chi$ of the order of
$\chi_\indexladder \sim 620$ for the two-leg ladders and
$\chi_\indexdimer \sim 2400$ for the dimers.

With the values of $\chi_\indexladder$ for the two-leg ladders, the
use of the standard scheme \cite{barthel_tdmrg_optimized_schemes,
  coira_temperature_dimer} of implementation of the time and temperature
evolution does not allow us to reach sufficiently long time and
resolution for the ladder case. This happens even though the dimers
and the two-leg ladders appear to be quite similar.  Thus, in order to be able to
study reliably the two-leg ladders we have implemented an optimal
numerical scheme as described in
Ref.~\onlinecite{barthel_tdmrg_optimized_schemes} (see
Fig.~\ref{fig:schemeC}), which was only scarcely used in the
literature previously due to its more demanding implementation.  As
shown in Fig.~\ref{fig:schemeC}, the usual procedure evolves  only one observable in
time, while the optimal scheme consists of
evolving both operators in time. In practice, this requires only
a logistic approach (storing of the state) and running the jobs in
parallel\cite{tange_gnuparallel}. However, due to hardware limits, it is in
practice difficult to store all the states and parallelize the
contraction properly.

We use the optimal scheme for the two-leg ladder case to increase the
resolution of Fig.~\ref{fig:channels_ladder025} and
Fig.~\ref{fig:channels_ladder059}. For the other models, we use the
normal scheme consisting of evolving only one observable in time
(Fig.~\ref{fig:schemeC} with $t_2=0$).

We compute the spin-spin correlations in space and time
\begin{equation}
 \average{S^\alpha_{x_0}(t) S^\gamma_{x_1}}_{\beta}
\end{equation}
for positive time $t \geq 0$ and fixed observable at $x_0$.  We detail
in Sec.~\ref{sec:channels} how we Fourier transform the measured
correlations.  We compute these correlations for various
temperatures as given in Table~\ref{table:temperature}.
\begin{table}
  \caption{ Temperatures presented in Figs.~\ref{fig:channels_ladder025}--\ref{fig:bosonization_ladder} and \ref{fig:channels_dimer_chain}. All the
    results are given at the \namemagn in the middle of the critical
    gapless phase. }
  \renewcommand{\arraystretch}{1.5}
  \begin{tabular*}{\columnwidth}{c@{\extracolsep{\fill}}cc}
    \hline\hline
    Model & Figure & Temperature  \\
    \hline
    ladder $\indexladder$ & Fig.~\ref{fig:channels_ladder025} & $T = 0.25~J_\perp \simeq 0.64~J_\parallel$ \\
          & Fig.~\ref{fig:channels_ladder059} & $T = 0.595~J_\perp \simeq 1.526~J_\parallel$ \\
     $\leftharpoonup$ \scriptsize{\textit{slices (Sec.\ref{sec:analytic})}}  $\rightarrow$   & Fig.~\ref{fig:bosonization_ladder} & $0.64~J_\parallel \lesssim T \lesssim 1.526~J_\parallel$ \\    
    \hline
    dimer $\indexdimer$ & Fig.~\ref{fig:channels_dimer025}  &  $T = 0.25~J_s \simeq 1.28~J_w/2$  \\
          & Fig.~\ref{fig:channels_dimer059}  &  $T = 0.595~J_s \simeq 3.05~J_w/2$ \\
          & Fig.~\ref{fig:channels_dimer_chain}  &  $T = 0.05~J_s \simeq 0.26~J_w/2$ \\
 $\leftharpoonup$ \scriptsize{$t^0$ \textit{excitation}}\;  $\rightarrow$   & Figs.~\ref{fig:channels_minimum},\ref{fig:channels_minimum_slices} & $0.05~J_s \lesssim T \lesssim 0.595~J_s$ \\    
 $\leftharpoonup$ \scriptsize{\textit{slices (Sec.\ref{sec:analytic})}}  $\rightarrow$   & Fig.~\ref{fig:bosonization_dimer} & $0.26~J_w/2 \lesssim T \lesssim 3.05~J_w/2$ \\    
    \hline
    chain $\indexchain$  & Fig.~\ref{fig:channels_chain025} &  $T = 0.25~J$ \\
    $\leftharpoonup$ \scriptsize{\textit{slices (Sec.\ref{sec:analytic})}}  $\rightarrow$   & Fig.~\ref{fig:bosonization_chain} & $0.1~J \lesssim T \lesssim 3.05~J$ \\    
    \hline\hline
  \end{tabular*}
  \label{table:temperature}
\end{table}

\subsection{Simulation accuracy}\label{sec:inputsim}

Comparing the initial \DMRG version with the finite temperature
algorithm, one sees that the variance of the Hamiltonian
$\average{\left(H-\average{H} \right)^2}$ is now finite and no longer
a criterion for convergence.  To define the convergence of the
calculation, we use the discarded weight
quantity\cite{barthel_tdmrg_optimized_schemes}, namely the total
weight of discarded eigenvalues according to the singular value
$\lambda_j$ decomposition.
\begin{equation}\label{eq:stop_criterion}
  \epsilon_i \equiv \left( \frac{\norm{X_{\scriptsize\text{trunc}}-X}}{\norm{X}} \right)^2 = \frac{\sum_{j>\chi} \lambda_{j}^2}{\sum_{j} \lambda_{j}^2}
\end{equation}
where $i$ can denote the inverse temperature $\beta$, the time $t$ or
a single step process depending on the context. The sum applies in all
quantum sector blocks. $X$ is in the matrix product
operator\cite{verstraete_garciaripoll_cirac_tdmrg_dissipative_breakthrough}
form and $X_{\scriptsize\text{trunc}}$ is the truncated matrix product
approximation with the renormalized bond dimension fixed to $\chi$.

The norm can be viewed as the Frobenius
norm\cite{barthel_tdmrg_optimized_schemes}.  $\chi$ is the bond
dimension of the mixed state, and corresponds to the number of states
kept in the system. This discarded weight quantity is a good indicator
of the \TDMRG algorithm precision (see Appendix~\ref{app:precision}).

\subsubsection{Ladders $\indexladder$}\label{sec:ladder_specific}

For the two-leg ladder, we fixed the size to a total of $2\cdot 45 = 90$
ladder sites. The run uses a truncation error
$\epsilon_\beta = 10^{-18}$ and steps in imaginary time
$\delta \beta=\frac{1}{100}$ to converge to the thermal equilibrium
$\beta=4.0~J_\perp^{-1}$, $3.04~J_\perp^{-1}$ and $1.68~J_\perp^{-1}$
which are the three temperatures considered for the ladders in the
present paper.

The initial bond dimension $\chi_\beta<800$ remains largely controlled
(it did not reach the limit size $800$) in this initial step since the
temperatures are quite large. Then we fix for all the different
observables the truncation $\epsilon_t = 10^{-13}$,
$\delta t=\frac{1}{16}$ and bond dimension $\chi_t = 620$.  Typically
the amount of information in the time simulation grows until the
maximal bond dimension is reached. Then, one loses a precision of
$\epsilon_i$ at each step by discarding the smallest singular values
$\lambda_j$ according to \eqref{eq:stop_criterion}, which is mandatory if
one wants to keep the numerical algorithmic complexity size $\chi$ of
the matrices fixed. The algorithm stops when the total discarded
weight passes a threshold
$\sum_{i \in \{\text{all steps}\}} \epsilon_i>10^{-2}$ or when a
single step lacks in precision $\epsilon_i>10^{-5}$.

The above precision is for the middle site observable (left column of
Fig.~\ref{fig:schemeC}). All the other observables run in parallel
with the same time algorithm procedure (or half the sites using
afterwards the symmetry along the ladder). In order to be sufficiently
fast, we reduce the bond dimension $\chi_t=400$ as well as the final
time $t_2 \sim 7-8~J_\perp^{-1}$ to ensure a precision
$\epsilon_t \lesssim 10^{-7}$. We can then compute the direct
correlations in Fig.~\ref{fig:schemeC} with $t_1$ up to
$ 19-23~J_\perp^{-1}$ (time within the bulk without encountered
borders) and we find a good overlap between all different time
correlations -- it gets a bit worse for values of $t_1$ close to the
maximal reachable time as expected. This optimal scheme brings an
increase in time $t=\abs{t_1}+\abs{t_2}$ or a resolution improvement of
$\sim 30 - 50 \%$ in the worst or best scenario.

\subsubsection{Dimers $\indexdimer$}\label{sec:dimer_specific}

For the dimer case, we get a similar resolution in $J_s$ without using
the optimal scheme for $L=90$ sites. We first converge with the
truncation error $\epsilon_\beta = 10^{-18}$ with steps
$\delta \beta = \frac{1}{100}$ to the thermal equilibrium at
$\beta=20~J_\perp^{-1}$, $10~J_\perp^{-1}$, $ 4.0~J_\perp^{-1}$,
$1.68~J_\perp^{-1}$. One then fixes all the different observables and
time evolves by $\delta t=\frac{1}{16}$ with the truncation
$\epsilon_t = 10^{-13}$ with the limited bond $\chi_t = 2400$. The
simulation stops again according to the same threshold and the final
times are for the high temperatures of order
$t_\text{max}\sim 27-58~J_\perp^{-1}$. For a lower temperature
$T \ll J_w$, one can achieve a better resolution similar to standard
\DMRG results (see Ref.~\onlinecite{coira_temperature_dimer}).

\section{Correlations for ladders, dimers and chains}\label{sec:channels}

\subsection{Dynamical structure factor}\label{sec:dyn_ss}

We first need to transform the real-time and -space data to find the
dynamical structure factor.
\begin{equation} \label{eq:dynamic_struct_factor}
 S^{\alpha \gamma}(\mathbf{q},\omega) = \int_{-\infty}^{\infty} \! \mathrm{d}\mathbf{r}\,\mathrm{d}t \,
 \mathrm{e}^{i(\omega t-\mathbf{q} \cdot \mathbf{r})}   \langle {S}^\alpha(\mathbf{r},t)   {S}^\gamma(0,0)  \rangle
\end{equation}
In the following equations, $\alpha, \gamma \in \{x,y,z,\pm \}$ denote
any component of the spin-$1/2$.

The \TDMRG simulation gives the direct correlations
$\average{S^{\alpha}_{x_0}(t)S^{\gamma}_{x_1}}_\beta$ restricted to
positive time $t \ge 0$. In order to avoid making errors using spatial
invariance too early (before time inversion which can be critical for
dimers) we use the retarded susceptibility to ensure an exact
procedure. Although this sounds \textit{a priori} more complicated, it actually
becomes more straightforward since the time symmetry is carried by the
Kramers-Kronig relations. Furthermore, the $S^z$ sector of the spin
does not need the average magnetization -- hidden in the real part --
to calculate the connected correlation.

\subsubsection{Chains $\indexchain$}\label{sec:dyn_chain}
We illustrate this procedure for the spin chain. We first Fourier
transform in time the retarded susceptibility that we get by
expressing it in a function of our direct correlations
\begin{align*}
  & \chi^{\alpha\gamma}_{\text{ret}}(\omega,x_0,x_1) = \int dt \; e^{+i (\omega + i \varepsilon) t} (-i\Theta(t)) \average{\left[ S^\alpha_{x_0}(t), S^\gamma_{x_1} \right] } \\
                                                &= -i \int_{0}^{+\infty} dt e^{+i (\omega + i \varepsilon) t} \left( \average{S^\alpha_{x_0}(t) S^\gamma_{x_1}} - \conj{\average{S^{\alpha\dagger}_{x_0}(t) S^{\gamma\dagger}_{x_1}}} \right)
\end{align*}
where $\conj{\average{\dots}}$ denotes complex number conjugation and
$S^{\pm\dagger}_{x_0}=S^{\mp}_{x_0}$. We then use translation
invariance $x=x_1-x_0$ in the bulk and Fourier transform the space
\begin{equation*}
  \chi^{\alpha\gamma}_{\text{ret}}(q,\omega) = \int d x \; e^{-i q x} \chi^{\alpha\gamma}_{\text{ret}}(\omega,x)
\end{equation*}
to finally get the dynamical structure factor worked out using the
Lehmann representation
\begin{equation*}
    S^{\alpha\gamma}(q,\omega) = \frac{-2}{1-e^{-\beta \omega}}\Im\left(\chi_\text{ret}^{\alpha\gamma}(q,\omega) \right)
\end{equation*}
The detail of this equality can be found in
Appendix~\ref{app:lehmann}.

\subsubsection{Ladders $\indexladder$}\label{sec:dyn_ladder}

For the ladder, we have two species of correlations according to the
leg index $\eta \in \{ 1,2 \}$. We use the $q_\perp$ momentum to
represent the correlations since it is a good quantum number. Thereby
the observables and correlations separate in the symmetric $q_\perp=0$
or antisymmetric $q_\perp=\pi$ sectors. We use the following
definitions
\begin{align*}
  S^\alpha_{\ell,q_\perp=0} &\equiv S^\alpha_{\ell,1} + S^\alpha_{\ell,2}  \\
  S^\alpha_{\ell,q_\perp=\pi} &\equiv S^\alpha_{\ell,1} - S^\alpha_{\ell,2}
\end{align*}
and calculate the dynamical structure factor from there. The two
correlations are fully represented in the two $q_\perp$ quantum
sectors
\begin{equation*}
 \average{S^\alpha_{q_\perp}(q,\omega)S^\gamma_{q_\perp} } = 2 \left( \average{S^\alpha_{1}(q,\omega)S^\gamma_{1} } \pm \average{S^\alpha_{1}(q,\omega)S^\gamma_{2} } \right)
\end{equation*}
where the index corresponds to correlations on the same legs or,
respectively, different legs. The symmetric $q_\perp=0$ case corresponds to the
sum while the antisymmetric $q_\perp=\pi$ case is the difference. Using the
rung symmetry, all other mixtures vanish,
$\average{S^\alpha_{q_\perp=0}(q,\omega)S^\gamma_{q_\perp=\pi} } =0$.

\subsubsection{Dimers $\indexdimer$}\label{sec:dyn_dimer}

Dimers have less symmetry than the two-leg ladder. One can map the dimer on
the ladder structure, but $q_\perp$ is not a good quantum number
anymore. This can be seen for the
$\average{S^{\alpha}_{\ell,1} S^{\gamma}_{0,2}}$ correlation, which is
not symmetric by space inversion anymore -- in contrast to the ladder case
(see Appendix~\ref{app:symmetries}). In one direction the correlation
starts with a weak bond while in the other direction it starts with a strong
bond. However, we can map the dimer on the ladder by introducing
similar definitions
\begin{align*}
  S^\alpha_{\ell,q_\perp=0} &\equiv ( S^\alpha_{\ell,1} + S^\alpha_{\ell,2} ) \\
  S^\alpha_{\ell,q_\perp=\pi} &\equiv ( S^\alpha_{\ell,1} - S^\alpha_{\ell,2} ) (-1)^{\ell}
\end{align*}
The new ladder labeling introduces the oscillating sign according to
Fig.~\ref{fig:models} while the symmetric observable
$S^\alpha_{\ell,q_\perp=0}$ remains untouched by the permutation. Note
that the crossed correlations
$\average{S^\alpha_{q_\perp=0}(q,\omega)S^\gamma_{q_\perp=\pi} } \neq
0$ are not vanishing but not very transparent to analyze. We thus
focus on the correlations
\begin{align*}
  \average{S^\alpha_{q_\perp=0}(q,\omega)S^\gamma_{q_\perp=0} } \\
  \average{S^\alpha_{q_\perp=\pi}(q,\omega)S^\gamma_{q_\perp=\pi} }
\end{align*}
These correlations look similar to the two-leg ladder up to finite
signal strength absent in the ladders as can be seen in
Fig.~\ref{fig:channels_dimer025} and Fig.~\ref{fig:channels_dimer059}.

For completeness, we also present in Appendix~\ref{app:along_dimer}
(Fig.~\ref{fig:channels_dimer_chain}) the results in the more
conventional chain notation in which we separate the correlations in
$\average{ S^\alpha_1(q,\omega) S^{\gamma}_1 }$ and
$\average{ S^\alpha_1(q,\omega) S^{\gamma}_2 }$, which are now two-site
cell translation invariant.

\subsubsection{Filter}\label{sec:filter}
In the previous paragraph, we made major assumptions such as
translation invariance and infinite time integration. Respectively, we
then should expect errors of the order
$\Delta q = \frac{\pi}{d_{max}}$ and $\Delta \omega = \frac{\pi}{t_{max}}$
inside each Fourier transformation due to finite-size effects.  As
usual, the most relevant error comes from the real-time resolution
which is much harder to get. In order to remove finite-size effects on
our data and give more weight to the short space and time steps, one
can introduce a selective mask.

In the first part
(Sec.~\ref{sec:results_correl}), we add a weak Gaussian filter
$M(x,t) = e^{-(A x / d_{max})^2} e^{-(B t / t_{max})^2}$ where we
choose $A=B=1.5$. 

In the second part (Sec.~\ref{sec:analytic}), where we compare the results
with the field theory expectation, we avoid all filters and work
with the raw correlations.

\begin{figure*}
  \centering
   \includegraphics[scale=0.8]{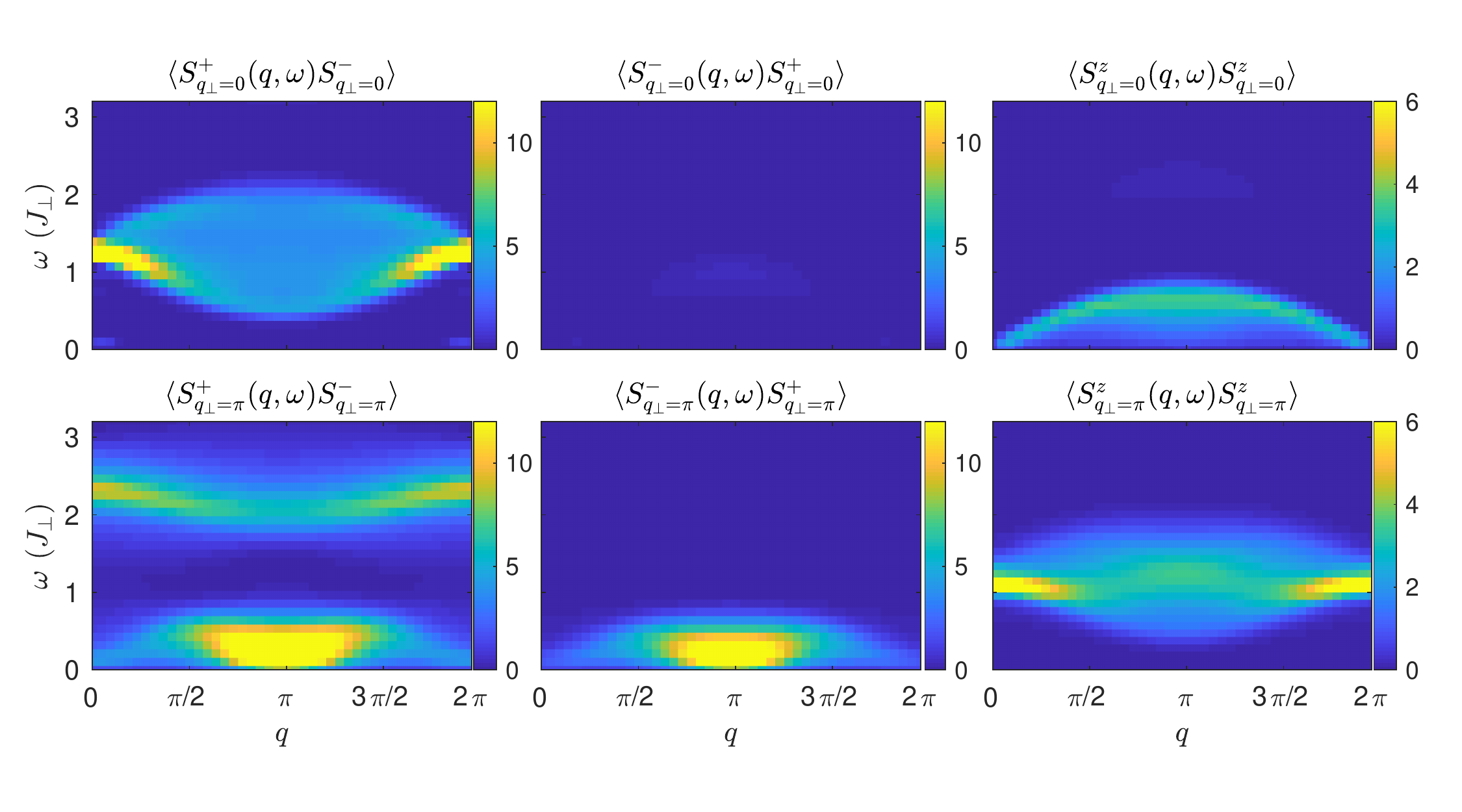}
   \caption{\label{fig:channels_ladder025} Spin-spin correlations of a
     spin-$1/2$ two-leg ladder in units of $J_\perp=1$ with
     $J_\parallel=0.39~J_\perp$ at
     $T=0.25~J_\perp \simeq 0.64~J_\parallel$ for magnetic field
     $h^z = 1.28~J_\perp$ corresponding to
     $m_{\indexladder}\simeq 0.5$. The resolution is
     $\Delta \omega = \frac{\pi}{30~J_\perp^{-1}}$ using the optimal
     scheme.}
\end{figure*}
\begin{figure*}
  \centering
   \includegraphics[scale=0.8]{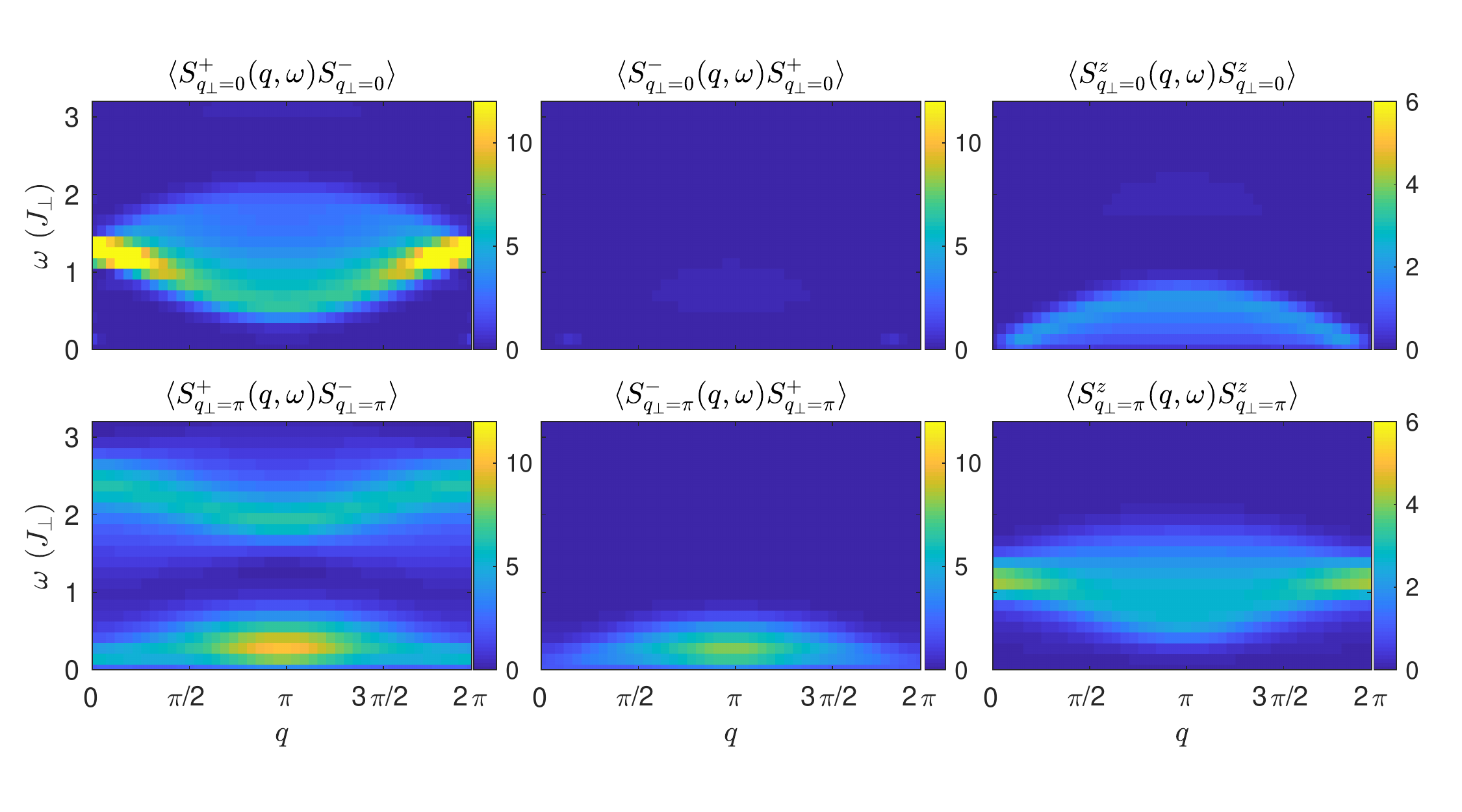}
   \caption{\label{fig:channels_ladder059} Spin-spin correlations of a
     spin-$1/2$ two-leg ladder in units of $J_\perp=1$ with
     $J_\parallel=0.39~J_\perp$ and at
     $T = 0.595~J_\perp \simeq 1.526~J_\parallel$ for magnetic
     field $h^z = 1.28~J_\perp$ corresponding to
     $m_{\indexladder}\simeq 0.5$. The resolution is
     $\Delta \omega = \frac{\pi}{22.5~J_\perp^{-1}}$ using the optimal
     scheme.}
\end{figure*}

\begin{figure*}
  \centering
   \includegraphics[scale=0.8]{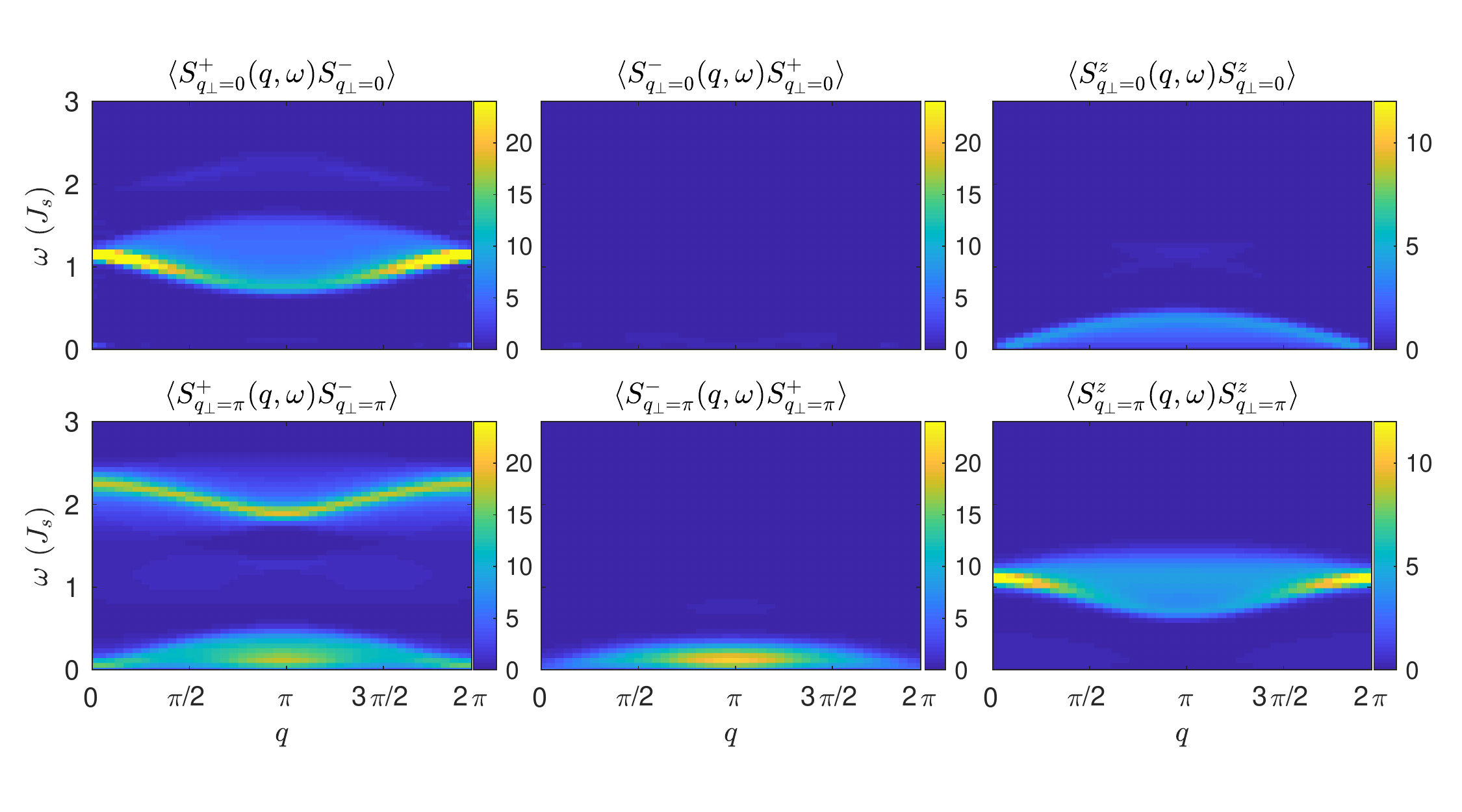}
   \caption{\label{fig:channels_dimer025} Weakly coupled dimerized chain in units of $J_s=1$ with
     $J_w=0.39~J_s$ at $T=0.25~J_s \simeq 1.28~J_w/2$ for magnetic field
     $h^z = 1.148~J_s$ corresponding to
     $m_{\indexdimer}\simeq 0.25$. The resolution is
     $\Delta \omega = \frac{\pi}{35~J_s^{-1}}$ using the standard
     scheme.}
\end{figure*}
\begin{figure*}
  \centering
   \includegraphics[scale=0.8]{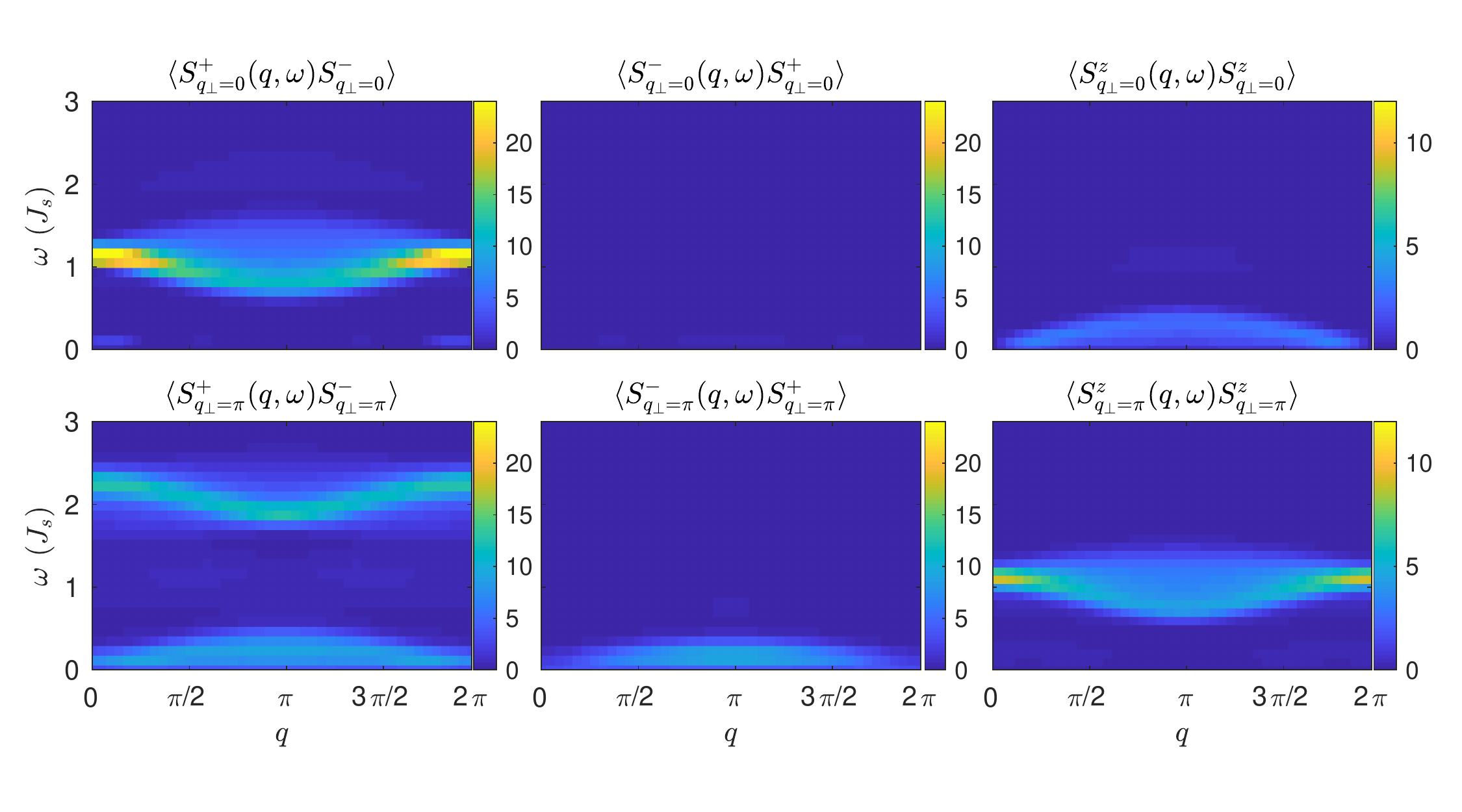}
   \caption{\label{fig:channels_dimer059} Weakly coupled dimerized
     chain in units of $J_s=1$ with $J_w= 0.39~J_s$ at
     $T = 0.595~J_s\simeq 3.05~J_w/2$ for magnetic field $h^z = 1.148~J_s$
     corresponding to $m_{\indexdimer} \simeq 0.25$. The resolution is
     $\Delta \omega = \frac{\pi}{27~J_s^{-1}}$ using the standard
     scheme.}
\end{figure*}

\subsection{Results for the correlation functions}\label{sec:results_correl}

Let us now present the results based on the calculations described in
the previous sections.

The central results of this section are the calculation of the
correlation functions for the ladders at finite temperature. Results
for the two-leg ladder model \eqref{eq:hamiltonian_ladder} are
presented in Fig.~\ref{fig:channels_ladder025} for a temperature of
$T=0.25~J_\perp$ and in Fig.~\ref{fig:channels_ladder059} for
temperature $T = 0.595~J_\perp$. Previous
results but at zero temperature $T=0$ can be found in
Ref.~\onlinecite{bouillot_long_ladder}.

\begin{figure*}
  \centering
   \includegraphics[scale=0.8]{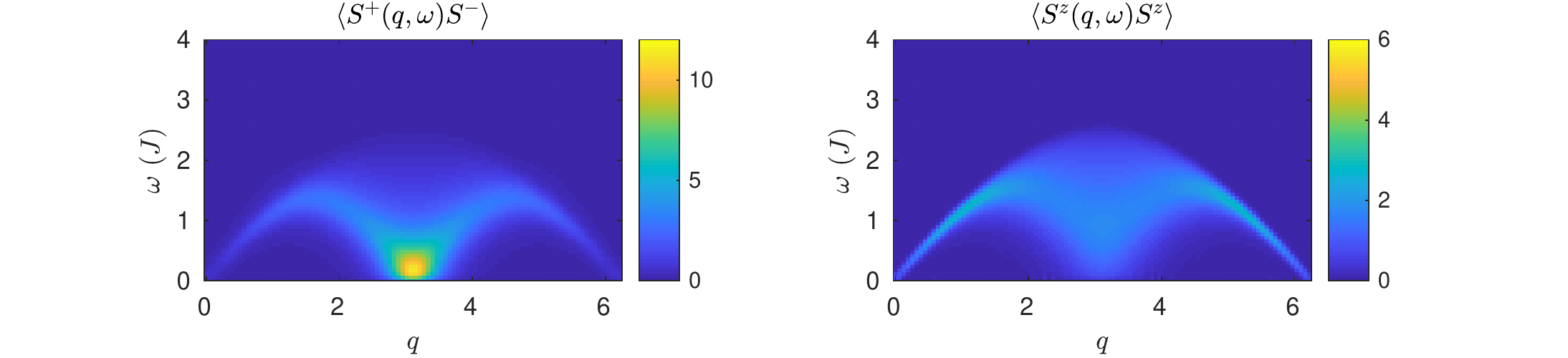}
   \caption{\label{fig:channels_chain025} \XXZ spin-$1/2$ chain with
     anisotropy $\Delta=\frac{1}{2}$ and coupling $J=1$ at $T=0.25~J$
     for an external magnetic field $h=0~J$ corresponding to a zero
     magnetization $m_{\indexchain}=0$. The resolution is
     $\Delta \omega = \frac{\pi}{52~J^{-1}}$.}
\end{figure*}

All allowed transitions for an isolated rung can be found in
Table~\ref{table:excitation_ladders} for the symmetric and
antisymmetric spin observables. Briefly we review the different
excitations that appear in the two-leg strong rung ladder, and we will
discuss it further in the next sections. In the gapless phase, the
lowest excitation spectrum is of course due to the interplay of the
singlets with the triplets forming the Descloizeaux-Pearson continuum
spectrum (see Sec.~\ref{sec:analytic} for a finer study of the low
spectrum behavior and Fig.~\ref{fig:channels_chain025}). At
intermediate energy, one sees the dispersion of a single triplet
excitation $\ket{t^0}$ in the correlations
$\average{S^{+-}_{q_\perp=0}}$ and $\average{S^{zz}_{q_\perp=\pi}}$
(see the $t-J$ model\cite{bouillot_long_ladder} which breaks down here
at finite $T\neq 0$, see Fig.~\ref{fig:channels_minimum}). In addition
to the same energy scale, there is a weak two-triplet excitation
signal in $\average{S^{-+}_{q_\perp=0}}$. At large energy scale, one
encounters another weak two-triplet excitation in the
$\average{S^{zz}_{q_\perp=0}}$ as well as a transition to the single
triplet $\ket{t^-}$ excitation in the
$\average{S^{+-}_{q_\perp=\pi}}$.

\begin{table}
  \caption{ Transition elements of the symmetric and antisymmetric
    spin-$1/2$ operator $\bra{\cdot} S^\alpha_{q_\perp} \ket{\cdot}$
    in the isolated rung picture. According to the Lehmann
    representation (see Appendix~\ref{app:lehmann}), all above-mentioned spectra can be identified by the following
    transitions. }
  \renewcommand{\arraystretch}{1.5}
  \begin{tabular*}{\columnwidth}{l|@{\extracolsep{\fill}}cccc}
    \hline\hline
     \scalebox{0.75}{$\mathbin{\rotatebox[origin=c]{55}{$\average{S^\alpha_{q_\perp}}$}}$} & $\ket{s}$ & $\ket{t^+}$ & $\ket{t^0}$ & $\ket{t^-}$  \\
    \hline
    $\bra{s}   $ & $\emptyset$        & $\scriptstyle\average{S^-_\pi}=-\sqrt{2}$ & $\scriptstyle\average{S^z_\pi}=1.0$ & $\scriptstyle\average{S^+_\pi}=\sqrt{2}$  \\
    $\bra{t^+} $ & $\scriptstyle\average{S^+_\pi}=-\sqrt{2}$ & $\scriptstyle\average{S^z_0 }=1.0$ & $\scriptstyle\average{S^+_0  }=\sqrt{2}$ & $\emptyset$         \\
    $\bra{t^0} $ & $\scriptstyle\average{S^z_\pi}=1.0$ & $\scriptstyle\average{S^-_0}=\sqrt{2}$  & $\emptyset$     & $\scriptstyle\average{S^+_0}=\sqrt{2}$  \\
    $\bra{t^-} $ & $\scriptstyle\average{S^-_\pi}=\sqrt{2}$ & $\emptyset$   & $\scriptstyle\average{S^-_0}=\sqrt{2}$   & $\scriptstyle\average{S^z_0}=-1$  \\ 
    \hline\hline
  \end{tabular*}
  \label{table:excitation_ladders}
\end{table}

In the same way, and in order to be able to compare with the ladder
results, we present the finite temperature correlations for the
dimerized system. Similar calculations, albeit at different
temperatures and couplings, were given in
Ref.~\onlinecite{coira_temperature_dimer}. The comparison of the results
of the present paper with the results of
Ref.~\onlinecite{coira_temperature_dimer} for the correlation
$\average{S_1^\alpha S_1^\gamma}_\beta$ in the gapless phase is very
good. We find similar limitations during the time evolution process for
the most cumbersome $\frac{e^{-\beta H}}{Z} S^+$ observable and
similar improvement of resolution when lowering the temperature.

The dimerized system mapped on the ladder geometry is shown in
Fig.~\ref{fig:channels_dimer025} at a temperature $T=0.25~J_s$ and in
Fig.~\ref{fig:channels_dimer059} for a temperature
$T = 0.595~J_s$. As one can see, most of the excitations can be
identified with the two-leg ladder pretty well.

Finally we also present the results for the chains at similar
temperatures in Fig.~\ref{fig:channels_chain025}. Finite temperature
calculations of spin chains were also presented in
Ref.~\onlinecite{barthel_xxz_Tdmrg}.

\subsection{Discussion of the \TDMRG results}\label{sec:discussion_results}

First let us note that the weights in the correlations redistribute
differently than for the zero-temperature
case\cite{bouillot_long_ladder}. Due to the finite temperature
effects, some negative energy transitions are allowed in the
correlations (see Fig.~\ref{fig:channels_dimer_chain}). We only
present the results for the positive frequency domain $\omega \geq 0$
since one can relate them to the negative frequencies using the
detailed balance equation \eqref{eq:detailed_balance}
\begin{equation}
  S^{\alpha \gamma}(q,\omega) = e^{-\beta \omega} S^{\gamma \alpha}(-q,-\omega)
\end{equation}
Since the raising and lowering are not self-conjugate operators, they
are allowed at finite temperature to get negative intensities (see
Fig.~\ref{fig:channels_dimer_chain}).

Even though the natures of the correlations are quite different due to
the different species of correlations, ladders and dimers are quite
related and give good information on each other. There are indeed
many similarities in the structure factors. The triplets are quite well
aligned in energy. One difference that can be directly seen in the
numerical results but will be deepened using the mapping onto an
effective spin chain (see Sec.~\ref{sec:comparison}) is that the dimer
has an effective dispersion of $J_w/2$ compared to $J_\parallel$ for
the two-leg ladder. All quantities depending on the weak bonds thus rescale
for the dimer by a factor of $2$. For this reason, we use a twice
larger colorbar color code for the dimer than the ladder to represent
the intensities on a similar scale -- but all presented results stay
in unity of $J_s$ and $J_\perp$.  In addition, due to the asymmetry of
the dimer, some signals survive in the region where the ladder is
actually gapped (correlations in $q_\perp=0$ and $q_\perp=\pi$ are not
allowed between the $\ket{t^+}$ and $\ket{t^0}$ or $\ket{t^0}$ and
$\ket{t^-}$). This is a remnant of the different types of geometries
and the absence of ``ladder-like'' symmetries in the dimer case (see
Appendix~\ref{app:symmetries}). In addition to that, the weak
two-triplet $\average{S^{-+}_{q_\perp=0}}$ at intermediate energy
scale is absent in the dimerized chain.

Concerning the effective temperature in each model, the two-leg ladder remains
more coherent in the low-energy spectrum than the dimer since it
``feels'' a twice smaller temperature in units of the effective
dispersion.  We discuss the low spectrum in more details in the next
section (Sec.~\ref{sec:analytic}) by comparing the results with field
theory predictions.

\begin{figure*}
  \centering
  \includegraphics[scale=0.825]{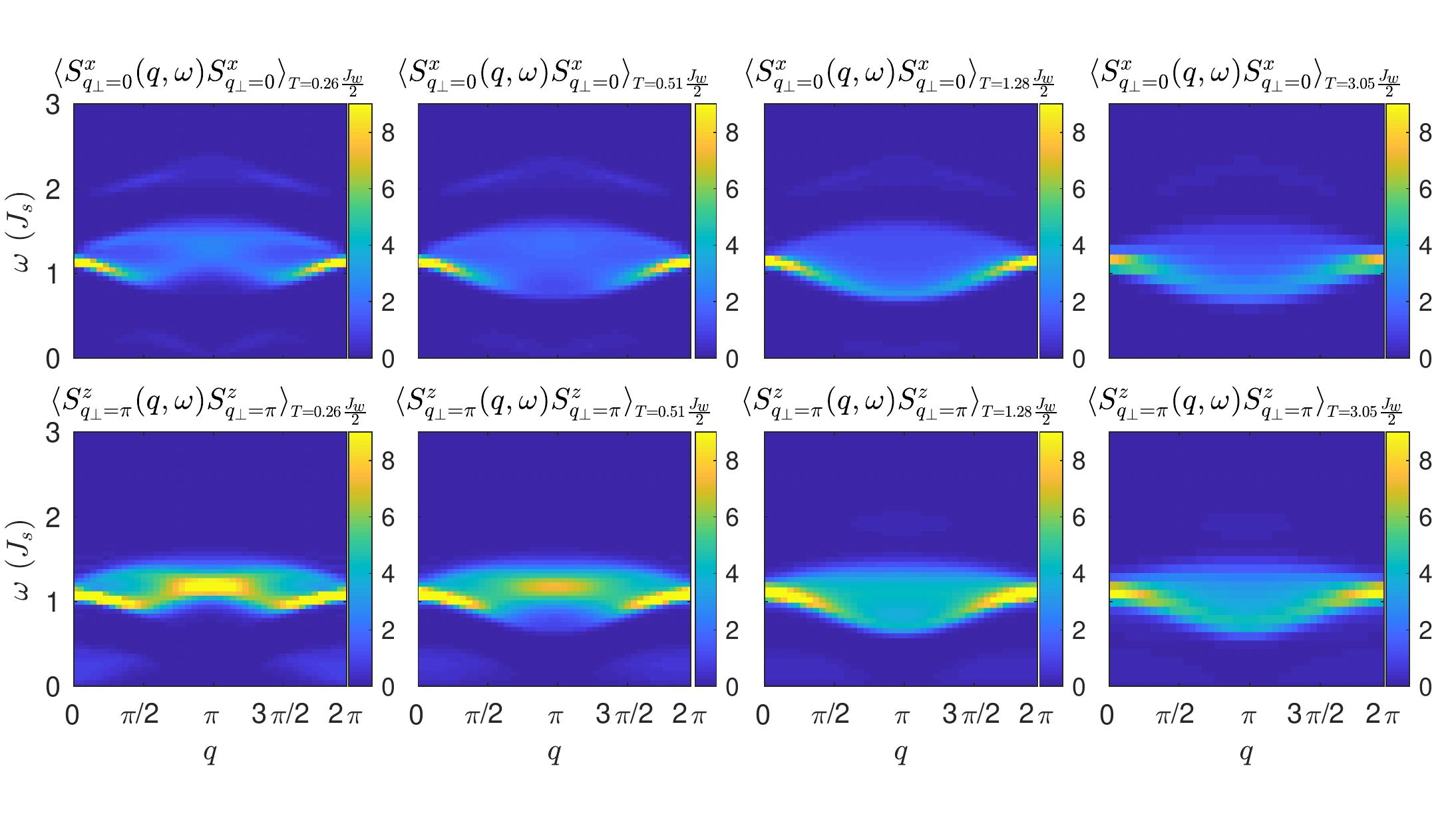}
  \caption{\label{fig:channels_minimum} The single triplet excitation
    $t^0$ in weakly coupled dimerized chain in units of $J_s=1$ with
    $J_w= 0.39~J_s$ at magnetic field $h^z = 1.148~J_s$ corresponding
    to $m_{\indexdimer} \simeq 0.25$ for various temperature
    $T = 0.05,\, 0.1 ,\, 0.25 ,\, 0.595~(J_s)$. The last two
    columns corresponds to top left and bottom right panels of
    Figs.~\ref{fig:channels_dimer025} and
    \ref{fig:channels_dimer059}. At the lowest temperatures we see
    that the minima of the dispersion are situated around $q=\pi/2$
    and $q=3\pi/2$ in agreement with the predictions of the mapping of
    this system to a $t-J$ model \cite{bouillot_long_ladder}.  When the
    temperature increases and becomes larger than $J_w$ one sees that
    a coherent-like mode with a minimum around $q=\pi$ appears at the
    bottom of the spectrum in an incoherent background (see text and
    Fig.~\ref{fig:channels_minimum_slices}).}
\end{figure*}

Let us now turn to the higher part of the energy spectrum.  This part
of the spectrum is of course beyond the reach of the field theory and
the mapping onto the anisotropic spin chain.  As a general tendency we
get weaker intensities and more spread signals when the temperature
increases. The temperature leads also to a broadening of the modes,
that was analyzed for the dimers from the numerical
results\cite{coira_temperature_dimer}.  We see here that the ladders
show similar behaviors in term of broadening (see
Figs.~\ref{fig:channels_ladder025} and \ref{fig:channels_ladder059}).
We concentrate here on the spectrum corresponding to an excitation to
the state $\ket{t^0}$ and study in detail its temperature dependence,
in particular for the same order or larger temperature than the weak
coupling (see Fig.~\ref{fig:channels_minimum}).  For both ladders and
dimers, this part of the spectrum corresponds to modes in which a
singlet or a $\ket{t^+}$ state is converted into a $\ket{t^0}$.  One
can thus examine this part of the spectrum as a single hole in a $t-J$
model \cite{bouillot_long_ladder} for which the ``hole'' corresponds
to the state $\ket{t^0}$ and the two ``spin'' states are played by the
singlet and $\ket{t^+}$ states. At low (or zero) temperature as was
clearly shown both for ladders at zero temperature (see Figs.~12,13 in
\onlinecite{bouillot_long_ladder}) and for dimers (see Figs.~11,12 in
\onlinecite{coira_temperature_dimer}) (note that the antisymmetric
signal $q_\perp = \pi$ is shifted by $\pi$ in our results compared to
the chain (in agreement with our
Fig.~\ref{fig:channels_dimer_chain})), the spectrum corresponds to two
cosine dispersions centered around the two minima $q=\pi/2$ and
$q=3\pi/2$ because the creation of a $\ket{t^0}$ state is accompanied
by the destruction of either a singlet $\ket{s}$ or a triplet
$\ket{t^+}$ (see Sec.~V.C.3.b in \onlinecite{bouillot_long_ladder}).
At the magnetic field we have applied, the low part of the excited
spectrum corresponds to the low-energy states $\ket{t^+}$ and
$\ket{s}$ with momentum for the excitations of $q=\pm \pi/2$ at half
filling.

From our numerical results we can follow the dispersion of the
$\ket{t^0}$ mode as the temperature increases from temperatures small
to large compared to the effective dispersion. The results are shown
in Fig.~\ref{fig:channels_minimum}.

\begin{figure*}
  \centering
  \includegraphics[scale=0.6]{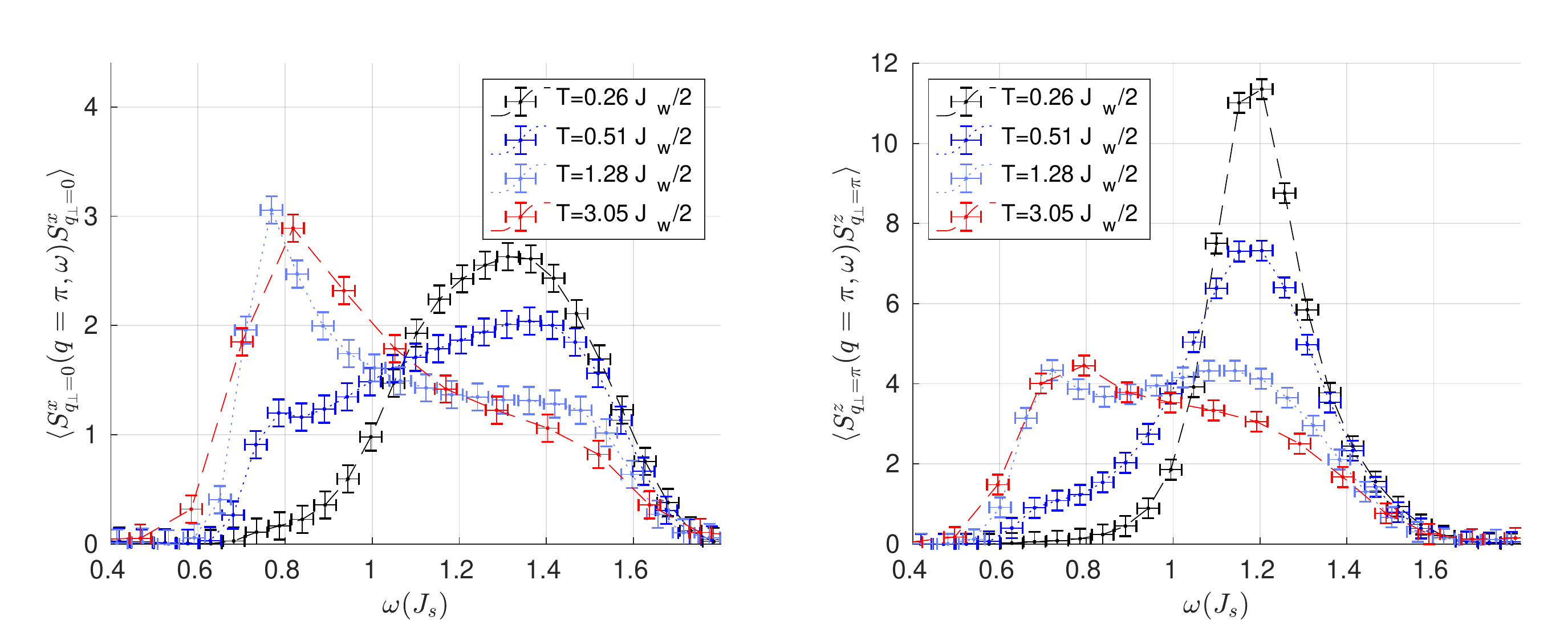}
  \caption{\label{fig:channels_minimum_slices} Slice at $q=\pi$ of
    Fig.~\ref{fig:channels_minimum} focusing on the energy range of
    the $\ket{t^0}$ excitation for the correlations
    $\average{S^x_{q_\perp=0}(q=\pi) S^x_{q_\perp=0}}$ on the left and
    $\average{S^z_{q_\perp=\pi}(q=\pi) S^z_{q_\perp=\pi}}$ on the
    right. We see a shift of the spectral weight there moving the mode
    from $1.2-1.4~J_s$ at low temperature to $0.7-0.85~J_s$ at large
    temperature. This complex mechanism appears to arise from the
    many-body dynamics and breaks the usual low-temperature picture
    done by the $t-J$ mapping\cite{bouillot_long_ladder}. }
\end{figure*}

The lower temperature is clearly in agreement with the previous
results both for the dimers and for the ladders with the dispersion minima
around $\pi/2$ and $3\pi/2$ resulting from the mapping to an effective
$t-J$ model. As the temperature increases we see that the modes become
increasingly incoherent and broaden. Quite surprisingly the numerical
result shows that the dispersion leads to a relevant intensity
corresponding to a coherent like mode, with its maximum intensity at the
bottom of the spectrum, with a minimum which is now shifted to around
$q\sim \pi$. This behavior is observed for the dimers as shown in
Figs.~\ref{fig:channels_minimum} and \ref{fig:channels_minimum_slices}, but also for ladders as can be
seen from Fig.~\ref{fig:channels_ladder059}.  Giving a precise
description of this effective ``mode'' is an interesting and challenging
question since it originally appears from the many-body
dynamics.
An interesting challenge would be finding the temperature for which the
incoherent minimal peak would become maximal.

Although it is difficult to connect this observation
directly to an analytical calculation, one can infer that the change
of the spectrum comes from the fact that the spinon excitations
that would correspond to the two pseudo-spin singlet $\ket{s}$
and triplet $\ket{t^+}$ states are now essentially totally incoherent since
the temperature is greater than their dispersion, leading to
essentially the dispersion of the bare hole.

\section{Comparison with field theory}\label{sec:analytic}

Let us now turn to the low-energy part of the spectra. Both the two-leg ladder
and the dimer system can be mapped\cite{bouillot_long_ladder,
  coira_temperature_dimer} at the \namemagn to an anisotropic
spin-$1/2$ $\Delta=\frac{1}{2}$ \XXZ model. This allows us to use the
standard bosonization method to extract the dynamical correlation
functions \cite{giam_1d_book} both at zero temperature, and using the
conformal invariance of the field theory, at low temperature.  In a
similar way to what was done for the NMR relaxation time
\cite{coira_giam_kollath_spin_lattice_relax_nmr_low_t} one can thus
compare the numerical results with the field theory description.

\subsection{Bosonization of the spin-$1/2$ chain}\label{sec:bosonization}

Let us give a brief reminder of the field theory description. One
introduces\cite{giam_1d_book} two continuous real bosonic fields
$\phi$ and $\theta$ to represent the low-energy excitations.  For an
\XXZ spin chain, the effective Hamiltonian is
\begin{equation}
  \label{eq:luttinger_quadratic}
  H = \frac{\hbar}{2\pi} \int\,dx \frac{u K}{\hbar^2} (\nabla \theta(x))^2 + \frac{u}{K} (\nabla \phi(x))^2
\end{equation}
where $u$ is the velocity of excitations and $K$ is a dimensionless
parameter, controlling the decay of the correlation functions. The
spin operators are represented in terms of the
fields $\phi$ and $\theta$ by\cite{giam_1d_book}
\begin{align}
  \label{eq:spin_phi_theta}
  S^z(\mathbf{r}) &= m_z + \frac{-1}{\pi} \nabla \phi(\mathbf{r}) + \frac{2}{2\pi \alpha} \cos(2\phi(\mathbf{r})-\pi(1+2m_z)x )  \nonumber \\
  S^\pm(\mathbf{r}) &= \frac{ e^{\mp i\theta(\mathbf{r})} }{\sqrt{2\pi \alpha}} \left( \cos(\pi x) + \cos(2\phi(\mathbf{r})-2 \pi m_z x ) \right)
\end{align}
where $m_z$ is the magnetization.

For ladders and dimerized chains, we use the spin-chain mapping
described in Sec.~\ref{sec:spinchainmapping} to relate the observables
to the ones of a spin chain.
\begin{equation}
\begin{split}
\pseudoS^z_l\; &= 2 S^z_{l,k} - \frac{1}{2} \\
 \pseudoS^\pm_l  &= (-1)^k \sqrt{2}~S^\pm_{l,k}
\end{split}
\end{equation}

The spin-spin correlation functions are given in the retarded
susceptibility form by\cite{schulz_quasi1D_sinegordon,
  chitra_giam_qcp_universal_function_ll, giam_1d_book}
\begin{widetext}
\begin{equation}
  \label{eq:bosonization_chi_ret}
    \chi_{\kappa}(\breve{q},\omega) =  - \frac{\sin(\pi \kappa)\alpha^2}{u} \left( \frac{2\pi\alpha}{\beta u} \right)^{2\kappa-2} \text{B}\left(\frac{\kappa}{2} - i \frac{\beta (\omega - u \breve{q} + i\varepsilon )}{4 \pi},1-\kappa\right) \text{B}\left(\frac{\kappa}{2}- i \frac{\beta (\omega + u \breve{q} +i\varepsilon )}{4 \pi},1-\kappa\right)
\end{equation}
\end{widetext}
where $\beta$ is the inverse temperature, $\alpha$ is a short
distance cutoff, and $\breve{q}$ is the momentum centered on the
field-dependent dispersion (the usual momentum $q$ is defined in
Sec.~\ref{sec:dyn_chain}). $\kappa$ is an exponent that depends on the
precise correlation function under consideration. In this paper we
look at the \namemagn and at slices at $q=\pi\pm 0.14$ (note that we
are slightly above half saturation too), which would correspond to the
$\breve{q}=0$ slice with \TLL exponents $2\kappa = 2 K $ or
$2 \kappa = \frac{1}{2K}$ according to equation \eqref{eq:slices}.

The non-universal parameters of the field theory (\TLL parameters and
amplitudes for the correlation functions) can be computed directly
allowing an essentially parameter free calculation of the correlation
functions. For the spin chain with $\Delta=\frac{1}{2}$, exact
Bethe-Ansatz results
\cite{giam_tsvelik_qcp_ladders_interplay_coupling} fix $K=0.75$ and
$u=1.299$. However, for the two-leg ladder and the dimer those parameters need
to be fixed from a numerical calculation with the microscopic model.

\begin{figure*}
  \centering
   \includegraphics[scale=0.5]{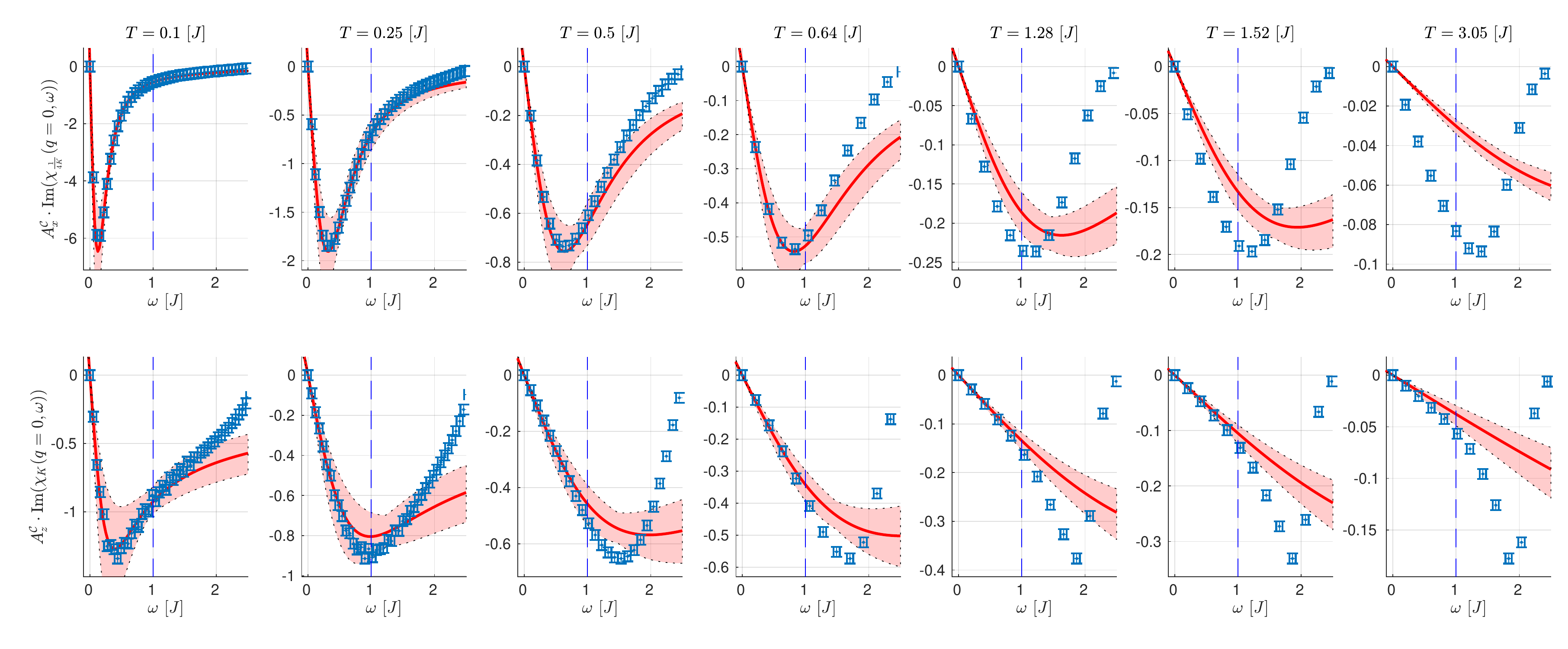}
   \caption{\label{fig:bosonization_chain} Transverse (top) and
     longitudinal (bottom) spin-spin correlation functions for a
     spin-$1/2$ chain as a function of the frequency $\omega$ for a
     fixed wave vector $q=\pi$ and $\breve{q}=0$, respectively (see
     Eq.~\eqref{eq:slices}). The field theory expression
     \eqref{eq:bosonization_chi_ret} in red is directly compared with
     the numerical \TDMRG calculation of the correlation (blue). The
     numerics are obtained from the Fourier transform of the output
     simulation without Gaussian filter (Sec.~\ref{sec:filter}). We
     have taken $A_x^\indexchain \simeq \CHAINAX$,
     $A_z^\indexchain \simeq \CHAINAZ$, and
     $K^\indexchain \simeq \CHAINK$. The shadow region corresponds to
     the maximum and minimum of all \TLL parameters moving by
     $\pm 10 \% $.}
\end{figure*}

\subsection{Extraction of \TLL parameters at $T=0$}\label{sec:parameters}

\begin{table}
  \caption{ \TLL parameters extracted from $T=0$ \DMRG data for the chain,
    dimer, and two-leg ladder systems. In italic in row $A_z$, $A^\indexchain_z$ could not
    be extracted directly from the \DMRG but has been fixed between
    the chain and the ladder (see text).}
  \renewcommand{\arraystretch}{1.5}
  \begin{tabular*}{\columnwidth}{c@{\extracolsep{\fill}}ccc}
  \hline\hline
  \TLL parameters & chain $\indexchain$ & dimer $\indexdimer$ & ladder $\indexladder$ \\
  \hline
  $A_x  $ & $\CHAINAX$ & $\DIMERAX$ & $\LADDERAX$ \\
  $B_x  $ & $\CHAINBX$ & $\DIMERBX$ & $\LADDERBX$ \\
  $A_z  $ & $\CHAINAZ$ & $\mathit{\DIMERAZ}$ & $\LADDERAZ$ \\
  $K  $ & $\CHAINK$ & $\DIMERK$ & $\LADDERK$ \\
  \hline\hline
  \end{tabular*}
  \label{table:LLparameter}
\end{table}

In order to fix the various parameters we use the expression of the
correlation functions at zero temperature\cite{giam_1d_book}
\begin{align*}
  \average{S^x_{i}S^x_{j}} &= (-1)^{|i-j|} A_x \left( \frac{1}{{|i-j|}} \right)^{\frac{1}{2K}} - B_x \left( \frac{1}{{|i-j|}} \right)^{2K+\frac{1}{2K}} \\
  \average{\delta S^z_{i}\delta S^z_{j}} &= \frac{-K}{2\pi^2} \left( \frac{1}{{|i-j|}} \right)^{2} + A_z (-1)^{|i-j|} \left( \frac{1}{{|i-j|}} \right)^{2K}
\end{align*}
These expressions can then be used, by comparison with the numerical
results, to extract \cite{hikihara_furusaki,bouillot_long_ladder} the
non-universal amplitudes $A_x$, $B_x$, and $A_z$ and the $K$ parameter. We
perform the zero-temperature \DMRG calculation of the correlation
functions using the ALPS library\cite{alps_mps}.

We first extracted the $A_x$ and $K$ parameter from the
$\average{S^x_iS^x_j}$ correlation since it has the slowest decay.  We
then use the obtained value of $K$ in the $\average{S^z_iS^z_j}$
correlation and fix $A_z$.  We avoid boundary effects by considering
correlations near the middle of the chain and by using space
invariance for few sites in the bulk. With this procedure, we estimate
all errors on the extracted values of about $20\%$. The velocity $u$
is computed from the compressibility
$\frac{K}{u\pi}=\frac{\partial m }{\partial h}$.  In this paper we used
$u\simeq 1.3$.  The \TLL values can be found in the table
\ref{table:LLparameter} and are consistent when they can be compared
with previous results\cite{hikihara_furusaki,bouillot_long_ladder}.

For the dimer system, it is more difficult than for the chain and the
ladder to extract the \TLL parameters.  For instance, the
$\average{S^z_i S^z_j}$ correlation decreases very fast while on the
other hand the local magnetization still oscillates. With the
asymmetry in the correlation, it becomes difficult to extract from
there any estimation of $A^\indexdimer_z$. This particular value has
thus been fixed to be between the chain and the ladder value.

\begin{figure*}
  \centering
   \includegraphics[scale=0.5]{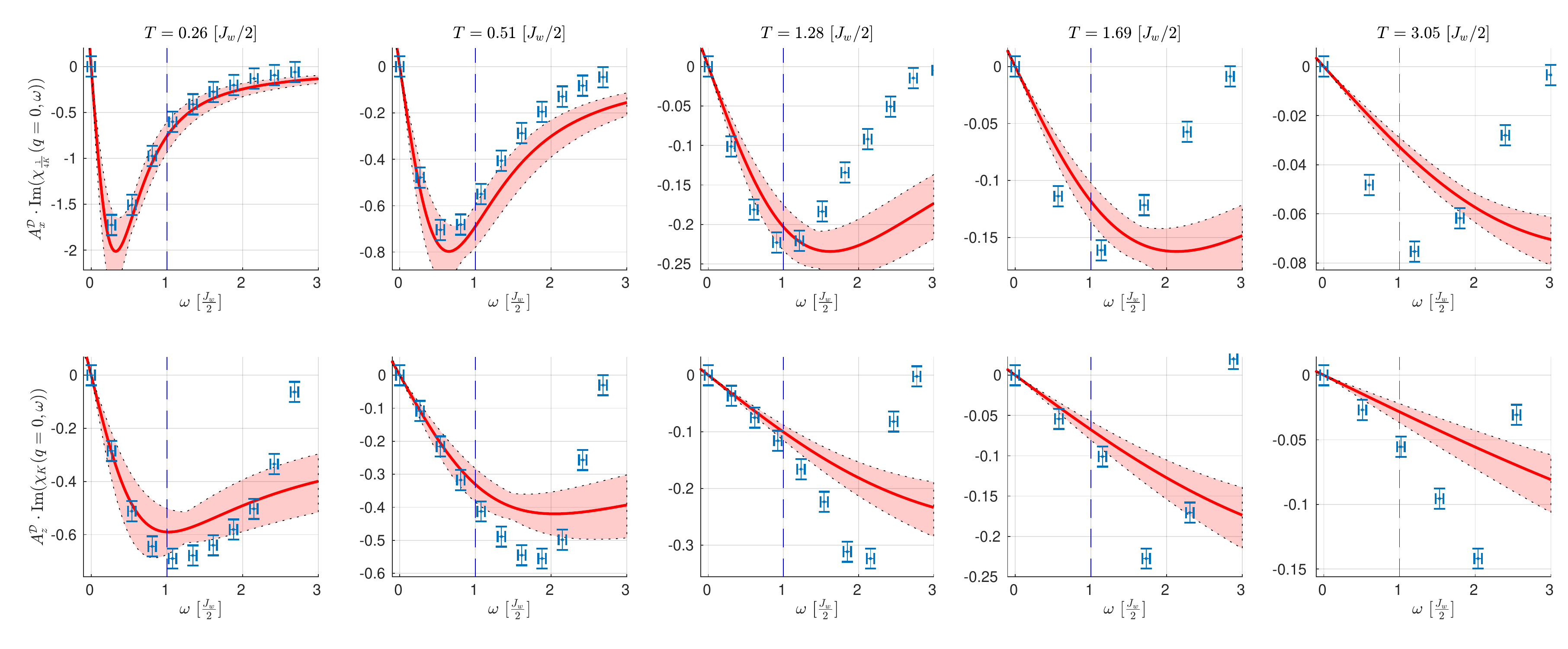}
   \caption{\label{fig:bosonization_dimer} Transverse (top) and
     longitudinal (bottom) spin-spin correlation functions for a
     dimerized chain as a function of the frequency $\omega$ for a
     fixed wave vector $q=\pi$ and $\breve{q}=0$, respectively (see
     Eq.~\eqref{eq:slices}). The field theory expression
     \eqref{eq:bosonization_chi_ret} in red is directly compared with
     the numerical \TDMRG calculation of the correlation (blue). The
     numerics are obtained from the Fourier transform of the output
     simulation without Gaussian filter (Sec.~\ref{sec:filter}). We have
     taken $A_x^\indexdimer \simeq \DIMERAX$,
     $A_z^\indexdimer \simeq \DIMERAZ$, and
     $K^\indexdimer \simeq \DIMERK$. The shadow region corresponds to
     the maximum and minimum of all \TLL parameters moving by
     $\pm 10 \% $.}
\end{figure*}

\begin{figure*}
  \centering
   \includegraphics[scale=0.5]{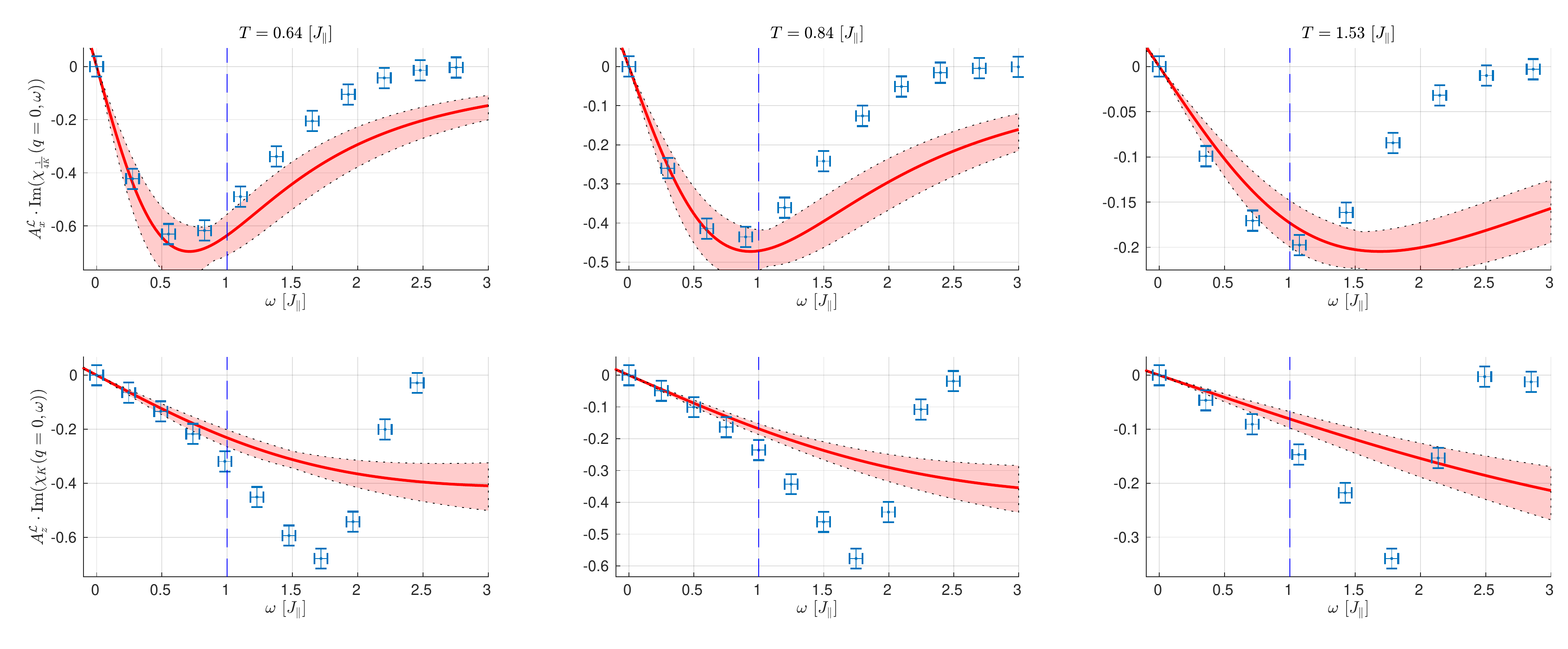}
   \caption{\label{fig:bosonization_ladder} Transverse (top) and
     longitudinal (bottom) spin-spin correlation functions for a
     two-leg ladder as a function of the frequency $\omega$ for a fixed
     wave vector $q=\pi$ and $\breve{q}=0$, respectively (see Eq.~\eqref{eq:slices}). The field theory expression
     \eqref{eq:bosonization_chi_ret} in red is directly compared with
     the numerical \TDMRG calculation of the correlation (blue). The
     numerics are obtained from the Fourier transform of the output
     simulation without Gaussian filter (Sec.~\ref{sec:filter}). We have
     taken $A_x^\indexladder \simeq \LADDERAX$,
     $A_z^\indexladder \simeq \LADDERAZ$, and
     $K^\indexladder \simeq \LADDERK$. The shadow region corresponds
     to the maximum and minimum of all \TLL parameters moving by
     $\pm 10 \% $.}
\end{figure*}

\subsection{Bosonization and \TDMRG comparison}\label{sec:comparison}

Since we have now fixed all the non-universal \TLL parameters and
amplitudes from Table~\ref{table:LLparameter}, we can use the field
theory expression (\ref{eq:bosonization_chi_ret}) to obtain the
correlation functions at finite temperature without any adjustable
parameter. For the comparison between the direct numerical calculation
of the correlations and the field theory, we consider the correlations
at $q=\pi$ which are directly related to \eqref{eq:swq_chi_prop} and
\eqref{eq:bosonization_chi_ret} by
\begin{equation}
  \label{eq:slices}
  \begin{split}
    A_x  \Im\left(  \chi_{\frac{1}{4K}}(\breve{q}=0,\omega)  \right) &= \frac{1-e^{-\beta \omega}}{-2} \average{\pseudoS^x(q=\pi,\omega) \pseudoS^x} \\
    A_z  \Im\left(  \chi_{K}(\breve{q}=0,\omega)  \right) &= \frac{1-e^{-\beta \omega}}{-2} \average{\delta \pseudoS^z(q=\pi,\omega) \delta \pseudoS^z}
  \end{split}
\end{equation}
The short-distance cutoff $\alpha$ can be taken as equal to $1$ inside
the retarded susceptibility since it is reabsorbed in the
non-universal amplitudes
$A_x = \left( \frac{\alpha^{\frac{1}{2K}-1}}{4\pi a^{\frac{1}{2K}-1} }
\right) $ and
$A_z=\left( \frac{\alpha^{2K-2}}{a^{2K-2} 2\pi^2} \right)$ according
to definition \eqref{eq:spin_phi_theta}, where $a$ is the lattice
spacing unit cell.

Let us first compare the field theory prediction with the numerical
calculations of the correlations for the anisotropic spin-$1/2$ chain.
The result is shown in Fig.~\ref{fig:bosonization_chain}.

As can be seen from the slices \ref{fig:bosonization_chain},
\ref{fig:bosonization_dimer}, and \ref{fig:bosonization_ladder} the
agreement is excellent both for the longitudinal and the transverse
correlations, for temperatures up to $T \lesssim 0.5 J$ for all
frequencies up to $\omega \lesssim J$ at which one would expect in any
case the field theory description to cease to be valid, irrespectively
of the thermal effects. Note that the frequency regime for which the
field theory is valid is much broader than what was the case for the
NMR relaxation time
\cite{coira_giam_kollath_spin_lattice_relax_nmr_low_t}.  This is
probably due to the fact that here we focus on a specific value of $q$
(slice) for which massless modes down to zero energy exist, rather
than perform a summation over all $q$ modes. It also confirms that for
a quite broad range of temperatures and frequencies, the conformal
modification of the zero-temperature correlations correctly gives the
finite temperature behavior. At larger temperatures $T > 0.64 J$ and
above, deviations start to appear, even if the low-energy part of the
spectrum remains remarkably robust even at quite high
temperatures. Note in particular the axis intensities in
Fig.~\ref{fig:bosonization_chain} that clearly show how well equation
\eqref{eq:bosonization_chi_ret} predicts the low spectrum behavior.

Quite remarkably, a similar excellent agreement is found for dimer and
ladder systems as, respectively, shown in
Fig.~\ref{fig:bosonization_dimer} and
Fig.~\ref{fig:bosonization_ladder}. The range of temperatures and frequencies for which the low-energy
effective theory works remarkably well is again quite broad.  Both
ladders and dimers also show an excellent agreement with the field
theory prediction for frequencies up to the natural cutoff of the
model, $J_\parallel$ for the two-leg ladder, or $J_w/2$ for the dimer
system. For the ladder, although we can only reach the relatively high
temperatures of more than half $J_\parallel$, the field theory remains
quite excellent up to frequencies of order
$\omega \lesssim J_\parallel$.

The extension of the TLL theory to finite temperature gives an
excellent quantitative description of the correlations up to
temperatures and energies close to the bandwidth of the problem.
This very robust behavior of the field theory description, in a broad
range of frequencies and temperatures, up to -- and sometimes even beyond
-- the natural cutoff of the theory is of course directly relevant in
the way that we can trust the application of such theories for treating
more complex realizations (such as, e.g., coupled systems).  This is of
course especially important to tackle the physics of compounds with
low enough magnetic exchanges, such that they can be manipulated by
realistic magnetic fields. The drawback of such compounds is of course
that the natural scale of energies (e.g., in a INS
experiment) or temperatures that one can reach is getting closer to
the magnetic exchange.

\section{Conclusion} \label{sec:conclusion}

In this paper, we computed using a \TDMRG technique the dynamical structure
factor of a two-leg spin-$1/2$ ladder system, as a function of the energy,
momentum, and temperature. We use an optimal scheme for the
implementation of the time evolution in order to be able to reach the
necessary resolution for the two-leg ladder system. We focus on the
intermediate magnetic field regime for which the magnetization per
rung or per dimer is half of the saturation value. There the system has
massless excitation and a low-energy part that can be mapped onto a
Tomonaga-Luttinger liquid.

The results are indicated in Fig.~\ref{fig:channels_ladder025} and
Fig.~\ref{fig:channels_ladder059}. We compare these spectra with those of dimerized systems and of an anisotropic $\Delta=\frac{1}{2}$ \XXZ
chain, to which the low-energy part of the previous systems can be
mapped. We examine in particular the evolution of the intermediate
energy part of the spectrum getting thermally populated by the triplet
$\ket{t^0}$. For the low-temperature part we examine the spin-chain
mapping and compare the finite temperature correlations with the
conformal modification of the \TLL field theory.  We show that there
is an excellent agreement between the numerics and the field theory
for energies and temperatures that extend up to values corresponding
to the spin exchange of the weak-coupling energy scale ($J_\parallel$
for ladders and $J_w/2$ for dimers).

Our paper shows clearly the direct possibility to use with an
excellent accuracy the field theory description to study more complex
systems of ladders such as weakly three-dimensional coupled ladders even if the
temperature or the interladder coupling reaches reasonably strong
values. It also shows that for systems as complex as the ladders we
have an essentially exact description even at finite temperatures from
the numerics and similar features can be found in related models (that
we have already analyzed in that way), namely, spin chains and
dimerized systems.

Our calculation can potentially be directly comparable to measurements
done with neutron scattering on two-leg ladder systems. Compounds such
as (C$_5$H$_{12}$N)$_2$CuBr$_4$ (\BPCB)\cite{thielemann_bpcb_neutrons_luttinger},
(C$_7$H$_{10}$N)$_2$CuBr$_4$ (\DIMPY)\cite{schmidiger_prl2013}, and (C$_5$H$_{12}$N)$_2$CuCl$_4$ (\BPCC)\cite{ward_bpcc_singlet} are
of course prime candidates for such study. Very successful comparisons
of the broad features of the neutrons have already been done with the
zero-temperature numerics and no high temperature as the one we have
studied yet exists in the literature in the gapless regime.  We hope
that the present paper will stimulate experimentalists to perform
these experiments, either in \BPCB or in similar compounds at larger
temperature $J_\parallel\lesssim T \lesssim J_\perp$, in particular to
probe the $\ket{t^0}$ incoherent dispersion.  For
\BPCB\cite{bouillot_long_ladder}, the couplings are, respectively,
$J_\parallel \simeq 3.55~\text{K}$ and $J_\perp \simeq 12.6~\text{K}$ and
the $\ket{t^0}$ mode is situated around the $J_\perp$ energy scale
(see Figs.~\ref{fig:channels_ladder025} and
\ref{fig:channels_ladder059}) and therefore located at a neutron energy of
approximately $1~\text{meV}$.  One could thus expect to see the
change of behavior for the $\ket{t^0}$ mode, as described by
Figs.~\ref{fig:channels_minimum} and
\ref{fig:channels_minimum_slices}, when going from $T=200~\text{mK}$
to $T=10~\text{K}$.

Our results open the door to a finer study of the temperature effects,
or the study via numerics of the vicinity of quantum critical points
in ladders for which such temperature effects are crucial to take into
account.

\begin{acknowledgments}
  We thank C. Berthod, P. Bouillot, N. A. Kamar, S. Takayoshi, S. Ward,
  and B. Wehinger for fruitful discussion. The calculations were done
  on the baobab and mafalda clusters at the University of Geneva. This
  work is supported in part by the Swiss National Science Foundation
  under Division II.
\end{acknowledgments}

\appendix

\section{\label{app:precision}Convergence and precision}

Note first that we use the Suzuki-Trotter decomposition in the
normalized units of the biggest energy scale to be consistent with the
diverse numerical precision and matrix conditioning.

\begin{figure}[h]
  \centering
  \includegraphics[scale=0.4]{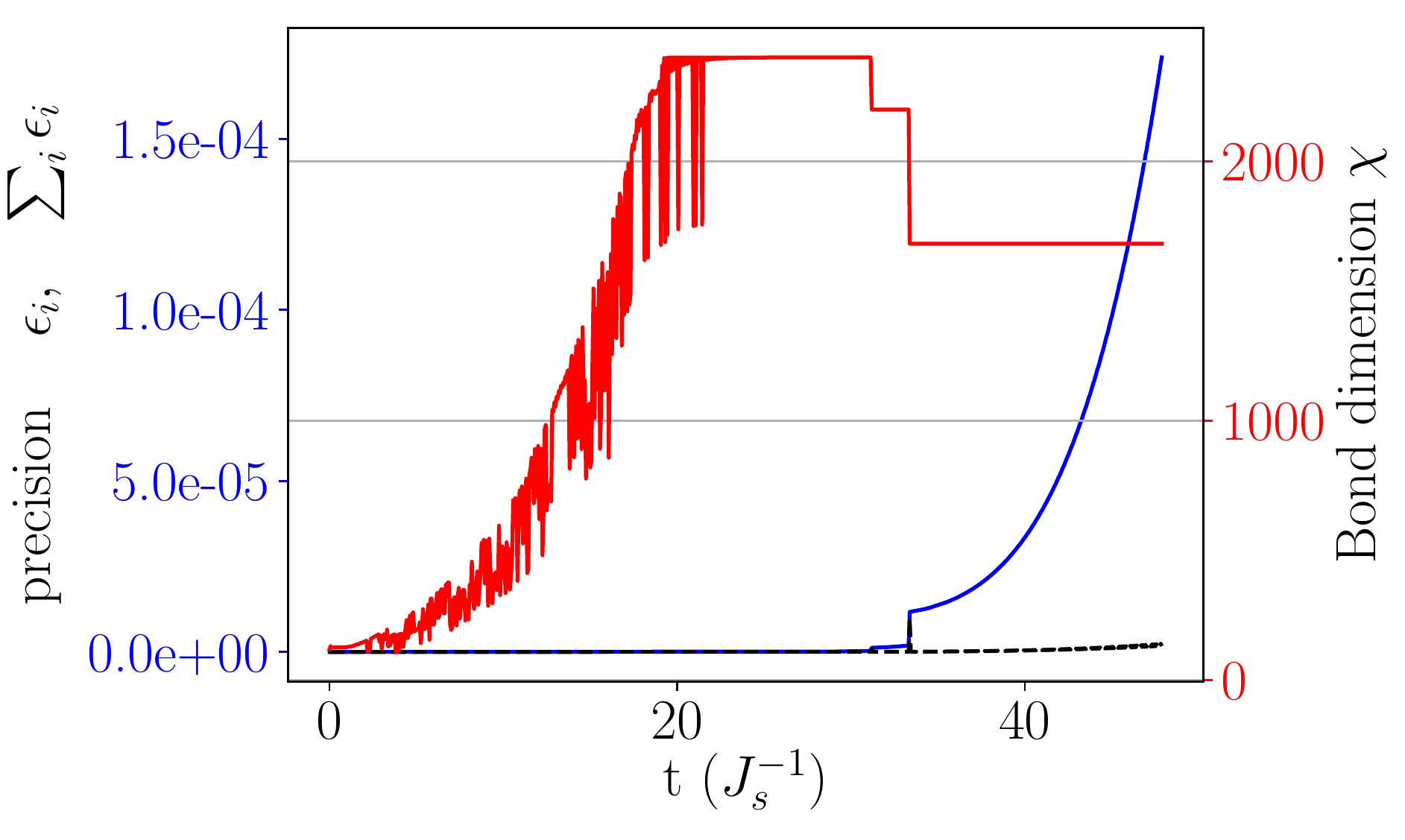}
  \includegraphics[scale=0.4]{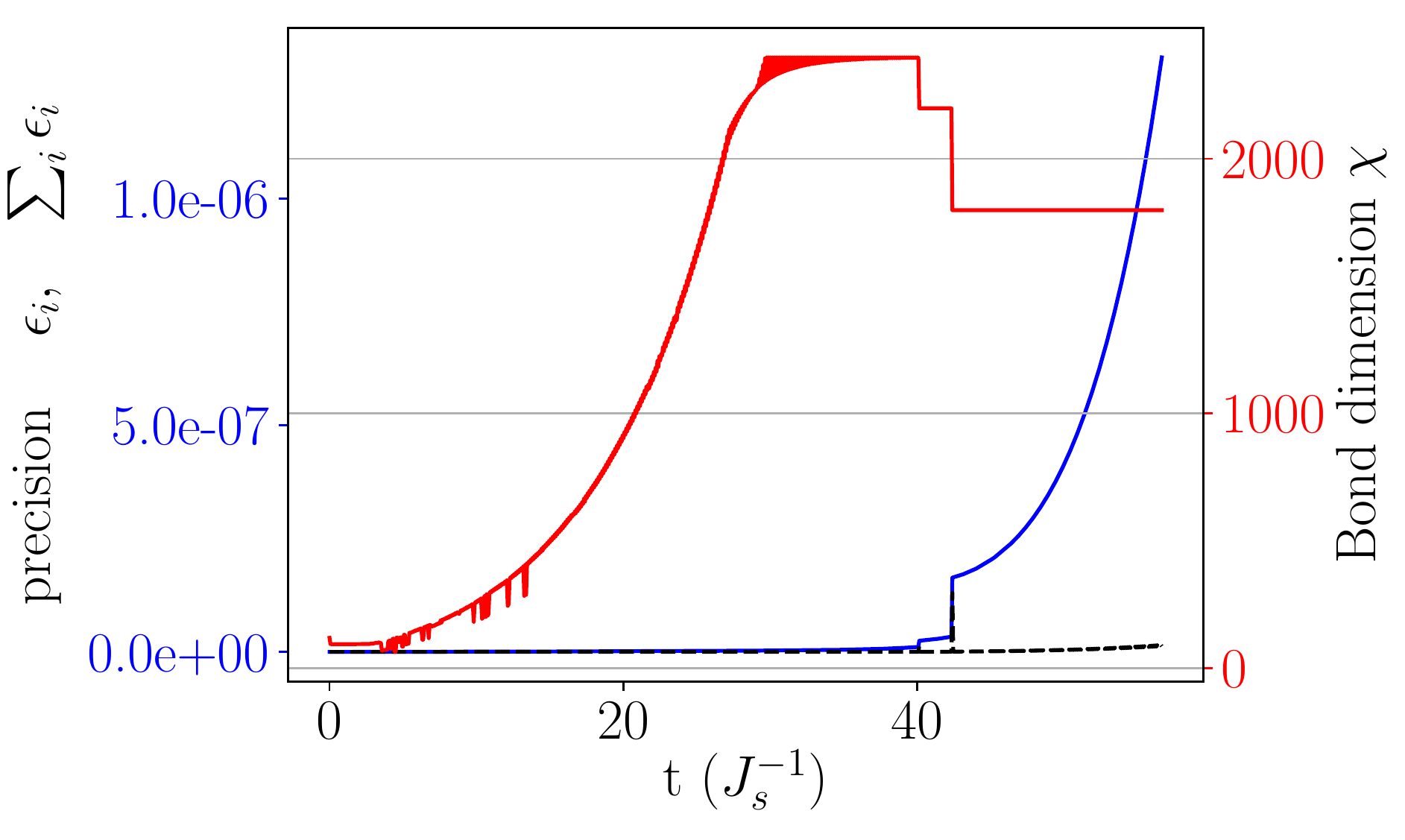}
  \includegraphics[scale=0.4]{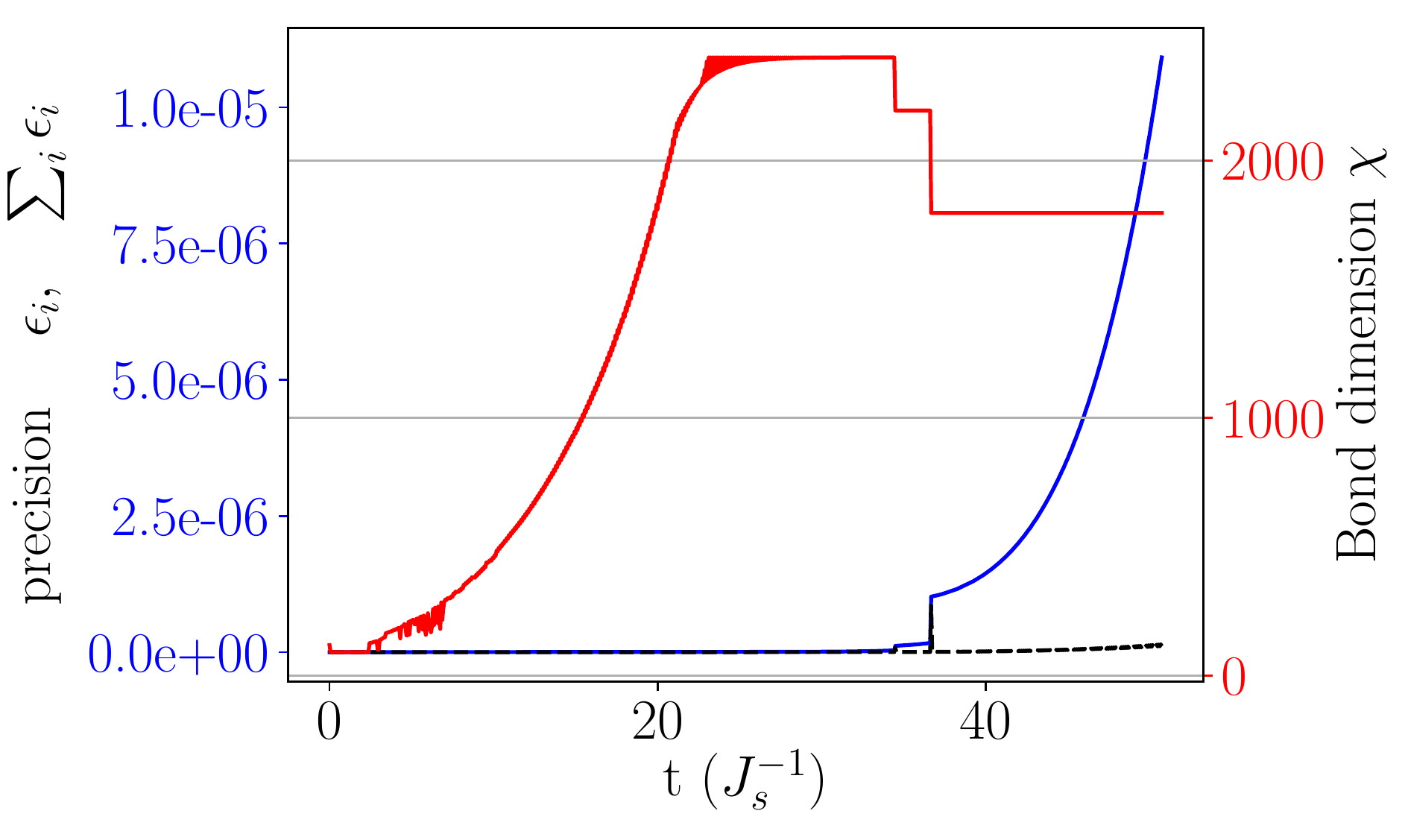}
  \caption{ Example of how the precision -- truncated weight
    $\epsilon_i$ (dashed black line), sum of all discarded weight
    $\sum_i \epsilon_i$ (solid blue line) and bond dimension $\chi$
    (red line) -- grows with the time for three different observables
    $e^{+iHt} \frac{e^{-\tilde{\beta} H}
      S^\alpha}{\sqrt{Z_{2\tilde{\beta}}}} e^{-iHt}$ in the ladder
    case at low temperature $\beta=2\tilde{\beta}=20.0~J_s^{-1}$. From
    top to bottom, we have $\alpha$ equal to ``$+$'', ``$-$'' and
    ``$z$''. The bond dimension truncation grows first until reaching
    $2400$. The computation starting to be too heavy, we reduce it to
    $2200$.  Then after a computational time of the order of a few
    days we reduced it again to $1700$ to reach reasonable
    computational times. }
  \label{fig:precision}
\end{figure}

As a rule of thumb, we set the maximal bond dimension for the problem
($\chi_\indexladder=620$, $\chi_\indexdimer=2400$ ). Of course, it
requires much less computational ressources to run the less consuming
observables ($\frac{e^{-\beta H}}{Z}S^-$ and
$\frac{e^{-\beta H}}{Z}S^z$, see Fig.~\ref{fig:precision}) at smaller
bond dimension $\chi$. However, the artificial oscillation would start
at different precision scales which we try to avoid. We always start
with some maximal value $\chi$ and then, if needed, reduce the bond
dimension to more quickly reach the final resolution of the problem. This
gives full accuracy for the initial short time evolution which reduces
the possibility of cumulative errors.

We present in Fig.~\ref{fig:precision} a plot of the bond dimension
with the truncated weight $\epsilon_i$ (one step) of the
Suzuki-Trotter process and the sum of all discarded weights
$\sum_i \epsilon_i$ (integration) for three observables.

The typically used measure of errors in the simulation, namely, the
$\epsilon_i$, is shown as the dashed black curve. In this paper we,
however, use the sum of all the discarded weights as the relevant error
$\sum_i \epsilon_i$. We believe that this more stringent criterion
helps to obtain results which are more accurate and reproducible.

\section{Lehmann representation and the detailed balance} \label{app:lehmann}

The Lehmann representation consists of computing the averages using
the exact eigenenergies $E_n$ and eigenvectors $|n\rangle$ of the
Hamiltonian:
\begin{equation*}
  \average{A(t)B} = \frac{1}{Z} \sum_{n,m} e^{-\beta E_n} e^{i(E_n-E_m)t} \bra{n} A \ket{m} \bra{m} B \ket{n}
\end{equation*}
It follows that the imaginary part of the susceptibility has the
following symmetries:
\begin{equation}
  \label{eq:chi_prop}
  \Im(\chi^{\alpha\gamma}_{\text{ret}}(q,\omega)) = - \Im(\chi^{\gamma\alpha}_{\text{ret}}(-q,-\omega))
\end{equation}
due to
\begin{align*}
  & { \scriptstyle \frac{-\pi}{Z} \sum_{n,m} (e^{-\beta E_n}-e^{-\beta E_m}) \bra{n} S^\alpha_{\frac{q}{2}} \ket{m} \bra{m} S^\gamma_{-\frac{q}{2}} \ket{n} \delta(\omega+E_n-E_m) = } \\
  & { \scriptstyle \frac{-\pi}{Z} \sum_{n,m} (e^{-\beta E_m}-e^{-\beta E_n})  \bra{n} S^\gamma_{-\frac{q}{2}} \ket{m} \bra{m} S^\alpha_{\frac{q}{2}} \ket{n} \delta(-\omega+E_n-E_m) }
\end{align*}
The dynamical structure factor is related to the imaginary
part of the susceptibility
\begin{equation}
  \label{eq:swq_chi_prop}
  S^{\alpha\gamma}(q, \omega) = \frac{-2}{1-e^{-\beta \omega}} \Im(\chi^{\alpha\gamma}_{\text{ret}}(q,\omega))
\end{equation}
due to
\begin{align*}
  & { \scriptstyle \frac{-\pi}{Z} \sum_{n,m} (e^{-\beta E_n}-e^{-\beta E_m}) \bra{n} S^\alpha \ket{m} \bra{m} S^\gamma \ket{n} \delta(\omega+E_n-E_m) } = \\
  & { \scriptstyle \frac{(1-e^{-\beta \omega})}{(-2)} \frac{(2\pi)}{Z} \sum_{n,m} e^{-\beta E_n}  \bra{n} S^\alpha \ket{m} \bra{m} S^\gamma \ket{n} \delta(\omega+E_n-E_m) }
\end{align*}

Thus the detailed balance equation follows from the two equations
\eqref{eq:chi_prop} and \eqref{eq:swq_chi_prop}
\begin{equation}
  \label{eq:detailed_balance}
  S^{\alpha \gamma}(q,\omega) = e^{-\beta \omega} S^{\gamma \alpha}(-q,-\omega)
\end{equation}

\section{Symmetries in ladders and dimers}\label{app:symmetries}

For the ladder, the rung and leg inversion symmetries for the middle
cell rung $\ell_0=\frac{N+1}{2}$ of a ladder with an odd number of rungs $N$
lead to
\begin{align*}
  \average{S^\alpha_{\eta_1,\ell}(t) S^\gamma_{\eta_2, \ell_0}} &=  \average{  S^\alpha_{\eta_2,-\ell}(t)  S^\gamma_{\eta_1, \ell_0} } = \\
  \average{S^\alpha_{\eta_2,\ell}(t) S^\gamma_{\eta_1, \ell_0}} &=  \average{  S^\alpha_{\eta_1,-\ell}(t)  S^\gamma_{\eta_2, \ell_0} }
\end{align*}
with $\eta_1, \eta_2 \in \{ 1, 2 \}$. All correlations are space
symmetric in the $\ell$ coordinate and we have equivalence between
top-top and bottom-bottom correlations as well as bottom-top and
top-bottom correlations. This makes the decomposition of the
correlation in the $q_\perp \in \{ 0, \pi \}$ sectors appropriate.

For the dimer, there is only one rung or leg symmetry. For the middle
rung cell $\ell_0 = \frac{N+1}{2}$ and for an odd number of rungs, we
have
\begin{align*}
  \average{ S^\alpha_{1,\ell}(t)  S^\gamma_{1, \ell_0} } &=  \average{  S^\alpha_{2,-\ell}(t)  S^\gamma_{2, \ell_0} } \neq \\
  \average{ S^\alpha_{2,\ell}(t)  S^\gamma_{2, \ell_0} } &= \average{  S^\alpha_{1,-\ell}(t)  S^\gamma_{1, \ell_0} }
\end{align*}
The left-left and right-right correlations have the same number of
couplings in both $\ell$ directions with reversed order
(strong$+$weak vs weak$+$strong). They are pretty similar up to
boundary effects.

The left-right and right-left correlations
\begin{align*}
  \average{ S^\alpha_{1,\ell}(t)  S^\gamma_{2, \ell_0} } &=  \average{  S^\alpha_{2,-\ell}(t)  S^\gamma_{1, \ell_0} } \neq \\
  \average{ S^\alpha_{2,\ell}(t)  S^\gamma_{1, \ell_0} } &=  \average{  S^\alpha_{1,-\ell}(t)  S^\gamma_{2, \ell_0} }
\end{align*}
are, however, very sensitive to the dimer geometry. For the first site
correlations $\ell-\ell_0\in\{-1,+1\}$, one crosses different amounts
of coupling in each direction (weak vs strong$+$weak). This asymmetry
makes those correlations very sensitive to the dimerization structure
even in the infinite-size limit. For those reasons, the
$q_\perp \in \{ 0, \pi \}$ is not a valid quantum number for the dimer
even though there exist many similarities with the ladder.

\section{Dimer spectrum along the chain direction}\label{app:along_dimer}

\begin{figure}[h]
  \centering
  \includegraphics[scale=0.5]{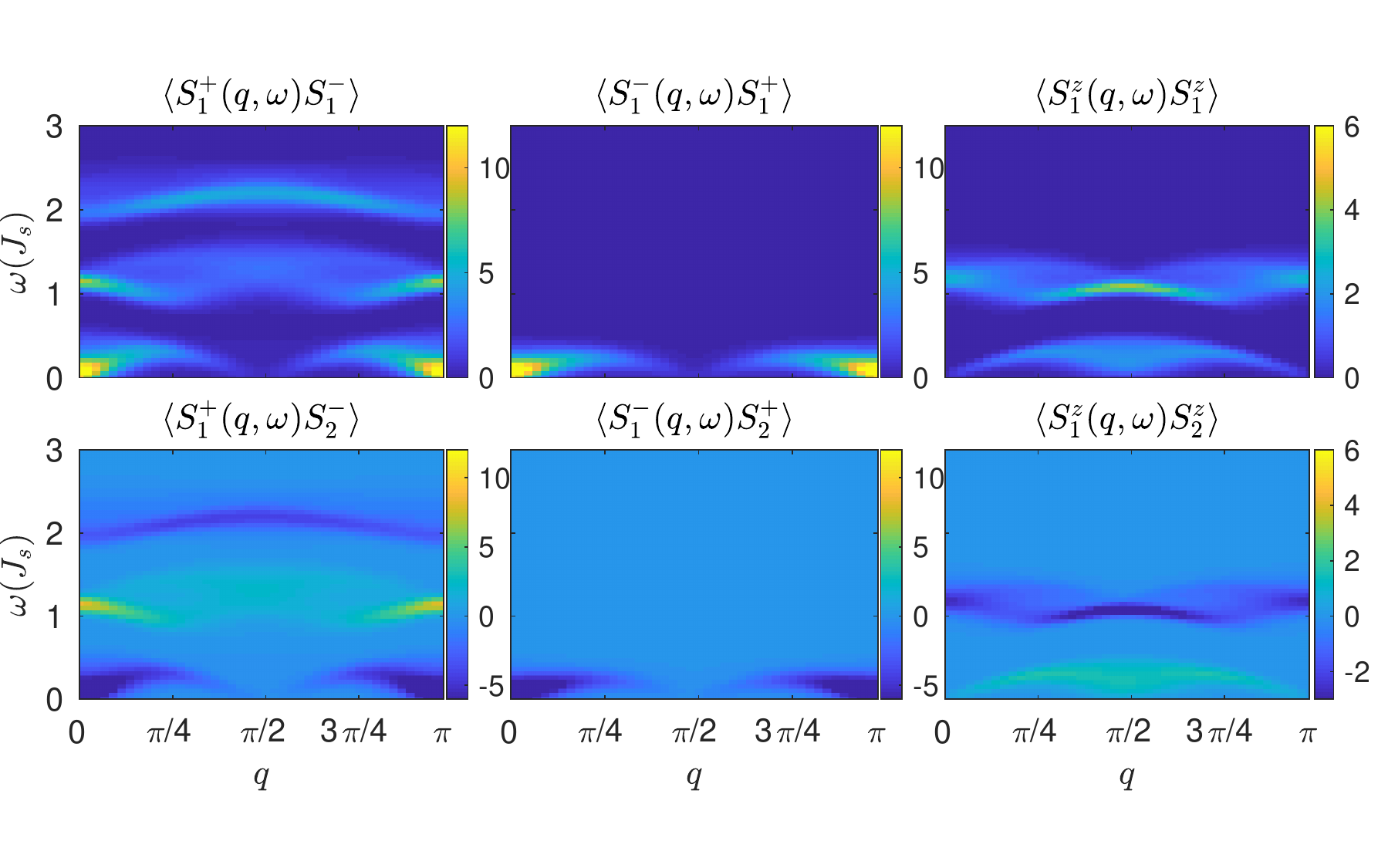}
  \caption{\label{fig:channels_dimer_chain} Correlations along the
    chain of the weakly coupled dimerized chain at $T=0.05~J_s$ with
    $J_w=0.39~J_s$ and $J_s=1$ for magnetic field corresponding to
    $m_\indexdimer \simeq 0.25$ (compare with
    Fig.~\ref{fig:channels_dimer025}). Note that, due to the
    geometry, the unit cells are now two sites periodic and the
    reciprocal space is $\pi$ periodic.}
\end{figure}

The main text presents the results of the dimer (see
Fig.~\ref{fig:channels_dimer025}) using the two-leg ladder
representation (as shown in Fig.~\ref{fig:models}).  For completeness
and more easy comparison with
Ref.~\onlinecite{coira_temperature_dimer} we also show in
Fig.~\ref{fig:channels_dimer_chain} the results in the chain
geometry.

\vspace{7cm}

The figure shows how the left-left cell
$\average{S^\alpha_1(q,\omega) S^\gamma_1}$ and left-right cell
$\average{S^\alpha_1(q,\omega) S^\gamma_2}$ disperse.


\begin{thebibliography}{53}
\makeatletter
\providecommand \@ifxundefined [1]{
 \@ifx{#1\undefined}
}
\providecommand \@ifnum [1]{
 \ifnum #1\expandafter \@firstoftwo
 \else \expandafter \@secondoftwo
 \fi
}
\providecommand \@ifx [1]{
 \ifx #1\expandafter \@firstoftwo
 \else \expandafter \@secondoftwo
 \fi
}
\providecommand \natexlab [1]{#1}
\providecommand \enquote  [1]{``#1''}
\providecommand \bibnamefont  [1]{#1}
\providecommand \bibfnamefont [1]{#1}
\providecommand \citenamefont [1]{#1}
\providecommand \href@noop [0]{\@secondoftwo}
\providecommand \href [0]{\begingroup \@sanitize@url \@href}
\providecommand \@href[1]{\@@startlink{#1}\@@href}
\providecommand \@@href[1]{\endgroup#1\@@endlink}
\providecommand \@sanitize@url [0]{\catcode `\\12\catcode `\$12\catcode
  `\&12\catcode `\#12\catcode `\^12\catcode `\_12\catcode `\%12\relax}%
\providecommand \@@startlink[1]{}
\providecommand \@@endlink[0]{}
\providecommand \url  [0]{\begingroup\@sanitize@url \@url }
\providecommand \@url [1]{\endgroup\@href {#1}{\urlprefix }}
\providecommand \urlprefix  [0]{URL }
\providecommand \Eprint [0]{\href }
\providecommand \doibase [0]{http://dx.doi.org/}
\providecommand \selectlanguage [0]{\@gobble}
\providecommand \bibinfo  [0]{\@secondoftwo}
\providecommand \bibfield  [0]{\@secondoftwo}
\providecommand \translation [1]{[#1]}
\providecommand \BibitemOpen [0]{}
\providecommand \bibitemStop [0]{}
\providecommand \bibitemNoStop [0]{.\EOS\space}
\providecommand \EOS [0]{\spacefactor3000\relax}
\providecommand \BibitemShut  [1]{\csname bibitem#1\endcsname}
\let\auto@bib@innerbib\@empty
\bibitem [{\citenamefont {Bloch}\ \emph {et~al.}(2008)\citenamefont {Bloch},
  \citenamefont {Dalibard},\ and\ \citenamefont {Zwerger}}]{bloch_MB_cold}
  \BibitemOpen
  \bibfield  {author} {\bibinfo {author} {\bibfnamefont {I.}~\bibnamefont
  {Bloch}}, \bibinfo {author} {\bibfnamefont {J.}~\bibnamefont {Dalibard}}, \
  and\ \bibinfo {author} {\bibfnamefont {W.}~\bibnamefont {Zwerger}},\ }\href
  {\doibase 10.1103/RevModPhys.80.885} {\bibfield  {journal} {\bibinfo
  {journal} {Reviews of Modern Physics}\ }\textbf {\bibinfo {volume} {80}},\
  \bibinfo {pages} {885} (\bibinfo {year} {2008})}\BibitemShut {NoStop}
\bibitem [{\citenamefont {Cazalilla}\ \emph {et~al.}(2011)\citenamefont
  {Cazalilla}, \citenamefont {Citro}, \citenamefont {Giamarchi}, \citenamefont
  {Orignac},\ and\ \citenamefont {Rigol}}]{cazalilla_condensedtocold_giam}
  \BibitemOpen
  \bibfield  {author} {\bibinfo {author} {\bibfnamefont {M.~A.}\ \bibnamefont
  {Cazalilla}}, \bibinfo {author} {\bibfnamefont {R.}~\bibnamefont {Citro}},
  \bibinfo {author} {\bibfnamefont {T.}~\bibnamefont {Giamarchi}}, \bibinfo
  {author} {\bibfnamefont {E.}~\bibnamefont {Orignac}}, \ and\ \bibinfo
  {author} {\bibfnamefont {M.}~\bibnamefont {Rigol}},\ }\href 
{\doibase  10.1103/RevModPhys.83.1405} {\bibfield  {journal} {\bibinfo  {journal}
  {Reviews of Modern Physics}\ }\textbf {\bibinfo {volume} {83}},\ \bibinfo
  {pages} {1405} (\bibinfo {year} {2011})}\BibitemShut {NoStop}
\bibitem [{\citenamefont {Auerbach}(1994)}]{assa_magnetism_book}
  \BibitemOpen
  \bibfield  {author} {\bibinfo {author} {\bibfnamefont {A.}~\bibnamefont
  {Auerbach}},\ }\href {http://www.springer.com/fr/book/9780387942865} {\emph
  {\bibinfo {title} {Interacting {Electrons} and {Quantum} {Magnetism}}}},\
  Graduate {Texts} in {Contemporary} {Physics}\ (\bibinfo  {publisher}
  {Springer-Verlag},\ \bibinfo {address} {New York},\ \bibinfo {year}
  {1994})\BibitemShut {NoStop}
\bibitem [{\citenamefont {Giamarchi}(2003)}]{giam_1d_book}
  \BibitemOpen
  \bibfield  {author} {\bibinfo {author} {\bibfnamefont {T.}~\bibnamefont
  {Giamarchi}},\ }\href@noop {} {\emph {\bibinfo {title} {Quantum Physics in
  One Dimension}}},\ International Series of Monographs on Physics\ (\bibinfo
  {publisher} {Oxford University Press},\ \bibinfo {address} {Oxford},\
  \bibinfo {year} {2003})\BibitemShut {NoStop}
\bibitem [{\citenamefont {Dagotto}\ and\ \citenamefont
  {Rice}(1996)}]{rice_quasi1vs2}
  \BibitemOpen
  \bibfield  {author} {\bibinfo {author} {\bibfnamefont {E.}~\bibnamefont
  {Dagotto}}\ and\ \bibinfo {author} {\bibfnamefont {T.~M.}\ \bibnamefont
  {Rice}},\ }\href {\doibase 10.1126/science.271.5249.618} {\bibfield
  {journal} {\bibinfo  {journal} {Science}\ }\textbf {\bibinfo {volume}
  {271}},\ \bibinfo {pages} {618} (\bibinfo {year} {1996})}\BibitemShut
  {NoStop}
\bibitem [{\citenamefont {Furrer}\ \emph {et~al.}(2009)\citenamefont {Furrer},
  \citenamefont {Mesot},\ and\ \citenamefont {Str\"assle}}]{ins_epfl_psi_book}
  \BibitemOpen
  \bibfield  {author} {\bibinfo {author} {\bibfnamefont {A.}~\bibnamefont
  {Furrer}}, \bibinfo {author} {\bibfnamefont {J.}~\bibnamefont {Mesot}}, \
  and\ \bibinfo {author} {\bibfnamefont {T.}~\bibnamefont {Str\"assle}},\
  }\href {\doibase 10.1142/4870} {\emph {\bibinfo {title} {Neutron Scattering
  in Condensed Matter Physics}}}\ (\bibinfo  {publisher} {WORLD SCIENTIFIC},\
  \bibinfo {year} {2009})\ \Eprint
  {http://arxiv.org/abs/https://www.worldscientific.com/doi/pdf/10.1142/4870}
  {https://www.worldscientific.com/doi/pdf/10.1142/4870} \BibitemShut {NoStop}
\bibitem [{\citenamefont {Berthier}\ \emph {et~al.}(2017)\citenamefont
  {Berthier}, \citenamefont {Horvati{\'c}}, \citenamefont {Julien},
  \citenamefont {Mayaffre},\ and\ \citenamefont
  {Kr{\"a}mer}}]{berthier_nuclear_2017}
  \BibitemOpen
  \bibfield  {author} {\bibinfo {author} {\bibfnamefont {C.}~\bibnamefont
  {Berthier}}, \bibinfo {author} {\bibfnamefont {M.}~\bibnamefont
  {Horvati{\'c}}}, \bibinfo {author} {\bibfnamefont {M.-H.}\ \bibnamefont
  {Julien}}, \bibinfo {author} {\bibfnamefont {H.}~\bibnamefont {Mayaffre}}, \
  and\ \bibinfo {author} {\bibfnamefont {S.}~\bibnamefont {Kr{\"a}mer}},\
  }\href {\doibase 10.1016/j.crhy.2017.09.009} {\bibfield  {journal} {\bibinfo
  {journal} {Comptes Rendus Physique}\ }\bibinfo {series} {2016 {Prizes} of the
  {French} {Academy} of {Sciences} /{Prix} 2016 de l'{Acad{\'e}mie} des
  sciences},\ \textbf {\bibinfo {volume} {18}},\ \bibinfo {pages} {331}
  (\bibinfo {year} {2017})}\BibitemShut {NoStop}
\bibitem [{\citenamefont {Caux}\ and\ \citenamefont
  {Calabrese}(2006)}]{caux_XXZ}
  \BibitemOpen
  \bibfield  {author} {\bibinfo {author} {\bibfnamefont {J.-S.}\ \bibnamefont
  {Caux}}\ and\ \bibinfo {author} {\bibfnamefont {P.}~\bibnamefont
  {Calabrese}},\ }\href {\doibase 10.1103/PhysRevA.74.031605} {\bibfield
  {journal} {\bibinfo  {journal} {Physical Review A}\ }\textbf {\bibinfo
  {volume} {74}},\ \bibinfo {pages} {031605} (\bibinfo {year}
  {2006})}\BibitemShut {NoStop}
\bibitem [{\citenamefont {Thielemann}\ \emph
  {et~al.}(2009{\natexlab{a}})\citenamefont {Thielemann}, \citenamefont
  {R{\"u}egg}, \citenamefont {R{\o}nnow}, \citenamefont {L{\"a}uchli},
  \citenamefont {Caux}, \citenamefont {Normand}, \citenamefont {Biner},
  \citenamefont {Kr{\"a}mer}, \citenamefont {G{\"u}del}, \citenamefont {Stahn},
  \citenamefont {Habicht}, \citenamefont {Kiefer}, \citenamefont {Boehm},
  \citenamefont {McMorrow},\ and\ \citenamefont
  {Mesot}}]{thielemann_spectrum_spinladder}
  \BibitemOpen
  \bibfield  {author} {\bibinfo {author} {\bibfnamefont {B.}~\bibnamefont
  {Thielemann}}, \bibinfo {author} {\bibfnamefont {C.}~\bibnamefont
  {R{\"u}egg}}, \bibinfo {author} {\bibfnamefont {H.~M.}\ \bibnamefont
  {R{\o}nnow}}, \bibinfo {author} {\bibfnamefont {A.~M.}\ \bibnamefont
  {L{\"a}uchli}}, \bibinfo {author} {\bibfnamefont {J.-S.}\ \bibnamefont
  {Caux}}, \bibinfo {author} {\bibfnamefont {B.}~\bibnamefont {Normand}},
  \bibinfo {author} {\bibfnamefont {D.}~\bibnamefont {Biner}}, \bibinfo
  {author} {\bibfnamefont {K.~W.}\ \bibnamefont {Kr{\"a}mer}}, \bibinfo
  {author} {\bibfnamefont {H.-U.}\ \bibnamefont {G{\"u}del}}, \bibinfo {author}
  {\bibfnamefont {J.}~\bibnamefont {Stahn}}, \bibinfo {author} {\bibfnamefont
  {K.}~\bibnamefont {Habicht}}, \bibinfo {author} {\bibfnamefont
  {K.}~\bibnamefont {Kiefer}}, \bibinfo {author} {\bibfnamefont
  {M.}~\bibnamefont {Boehm}}, \bibinfo {author} {\bibfnamefont {D.~F.}\
  \bibnamefont {McMorrow}}, \ and\ \bibinfo {author} {\bibfnamefont
  {J.}~\bibnamefont {Mesot}},\ }\href {\doibase 10.1103/PhysRevLett.102.107204}
  {\bibfield  {journal} {\bibinfo  {journal} {Physical Review Letters}\
  }\textbf {\bibinfo {volume} {102}},\ \bibinfo {pages} {107204} (\bibinfo
  {year} {2009}{\natexlab{a}})}\BibitemShut {NoStop}
\bibitem [{\citenamefont {Thielemann}\ \emph
  {et~al.}(2009{\natexlab{b}})\citenamefont {Thielemann}, \citenamefont
  {R{\"u}egg}, \citenamefont {Kiefer}, \citenamefont {R{\o}nnow}, \citenamefont
  {Normand}, \citenamefont {Bouillot}, \citenamefont {Kollath}, \citenamefont
  {Orignac}, \citenamefont {Citro}, \citenamefont {Giamarchi}, \citenamefont
  {L{\"a}uchli}, \citenamefont {Biner}, \citenamefont {Kr{\"a}mer},
  \citenamefont {Wolff-Fabris}, \citenamefont {Zapf}, \citenamefont {Jaime},
  \citenamefont {Stahn}, \citenamefont {Christensen}, \citenamefont {Grenier},
  \citenamefont {McMorrow},\ and\ \citenamefont
  {Mesot}}]{thielemann_bpcp_neutron_ladder}
  \BibitemOpen
  \bibfield  {author} {\bibinfo {author} {\bibfnamefont {B.}~\bibnamefont
  {Thielemann}}, \bibinfo {author} {\bibfnamefont {C.}~\bibnamefont
  {R{\"u}egg}}, \bibinfo {author} {\bibfnamefont {K.}~\bibnamefont {Kiefer}},
  \bibinfo {author} {\bibfnamefont {H.~M.}\ \bibnamefont {R{\o}nnow}}, \bibinfo
  {author} {\bibfnamefont {B.}~\bibnamefont {Normand}}, \bibinfo {author}
  {\bibfnamefont {P.}~\bibnamefont {Bouillot}}, \bibinfo {author}
  {\bibfnamefont {C.}~\bibnamefont {Kollath}}, \bibinfo {author} {\bibfnamefont
  {E.}~\bibnamefont {Orignac}}, \bibinfo {author} {\bibfnamefont
  {R.}~\bibnamefont {Citro}}, \bibinfo {author} {\bibfnamefont
  {T.}~\bibnamefont {Giamarchi}}, \bibinfo {author} {\bibfnamefont {A.~M.}\
  \bibnamefont {L{\"a}uchli}}, \bibinfo {author} {\bibfnamefont
  {D.}~\bibnamefont {Biner}}, \bibinfo {author} {\bibfnamefont {K.~W.}\
  \bibnamefont {Kr{\"a}mer}}, \bibinfo {author} {\bibfnamefont
  {F.}~\bibnamefont {Wolff-Fabris}}, \bibinfo {author} {\bibfnamefont {V.~S.}\
  \bibnamefont {Zapf}}, \bibinfo {author} {\bibfnamefont {M.}~\bibnamefont
  {Jaime}}, \bibinfo {author} {\bibfnamefont {J.}~\bibnamefont {Stahn}},
  \bibinfo {author} {\bibfnamefont {N.~B.}\ \bibnamefont {Christensen}},
  \bibinfo {author} {\bibfnamefont {B.}~\bibnamefont {Grenier}}, \bibinfo
  {author} {\bibfnamefont {D.~F.}\ \bibnamefont {McMorrow}}, \ and\ \bibinfo
  {author} {\bibfnamefont {J.}~\bibnamefont {Mesot}},\ }\href 
{\doibase  10.1103/PhysRevB.79.020408} {\bibfield  {journal} {\bibinfo  {journal}
  {Physical Review B}\ }\textbf {\bibinfo {volume} {79}},\ \bibinfo {pages}
  {020408} (\bibinfo {year} {2009}{\natexlab{b}})}\BibitemShut {NoStop}
\bibitem [{\citenamefont {White}(1992)}]{white_dmrg_article1_breakthrough}
  \BibitemOpen
  \bibfield  {author} {\bibinfo {author} {\bibfnamefont {S.~R.}\ \bibnamefont
  {White}},\ }\href {\doibase 10.1103/PhysRevLett.69.2863} {\bibfield
  {journal} {\bibinfo  {journal} {Phys. Rev. Lett.}\ }\textbf {\bibinfo
  {volume} {69}},\ \bibinfo {pages} {2863} (\bibinfo {year}
  {1992})}\BibitemShut {NoStop}
\bibitem [{\citenamefont {White}(1993)}]{white_dmrg_article2_breakthrough}
  \BibitemOpen
  \bibfield  {author} {\bibinfo {author} {\bibfnamefont {S.~R.}\ \bibnamefont
  {White}},\ }\href {\doibase 10.1103/PhysRevB.48.10345} {\bibfield  {journal}
  {\bibinfo  {journal} {Phys. Rev. B}\ }\textbf {\bibinfo {volume} {48}},\
  \bibinfo {pages} {10345} (\bibinfo {year} {1993})}\BibitemShut {NoStop}
\bibitem [{\citenamefont
  {Vidal}(2003)}]{vidal_tebd_precursor_pure_state_dynamics_breakthrough}
  \BibitemOpen
  \bibfield  {author} {\bibinfo {author} {\bibfnamefont {G.}~\bibnamefont
  {Vidal}},\ }\href {\doibase 10.1103/PhysRevLett.91.147902} {\bibfield
  {journal} {\bibinfo  {journal} {Phys. Rev. Lett.}\ }\textbf {\bibinfo
  {volume} {91}},\ \bibinfo {pages} {147902} (\bibinfo {year}
  {2003})}\BibitemShut {NoStop}
\bibitem [{\citenamefont {Daley}\ \emph {et~al.}(2004)\citenamefont {Daley},
  \citenamefont {Kollath}, \citenamefont {Schollw{\"o}ck},\ and\ \citenamefont
  {Vidal}}]{daley_kollath_schollwock_tdmrg_pure_state_dynamics_breakthrough}
  \BibitemOpen
  \bibfield  {author} {\bibinfo {author} {\bibfnamefont {A.~J.}\ \bibnamefont
  {Daley}}, \bibinfo {author} {\bibfnamefont {C.}~\bibnamefont {Kollath}},
  \bibinfo {author} {\bibfnamefont {U.}~\bibnamefont {Schollw{\"o}ck}}, \ and\
  \bibinfo {author} {\bibfnamefont {G.}~\bibnamefont {Vidal}},\ }\href
  {http://stacks.iop.org/1742-5468/2004/i=04/a=P04005} {\bibfield  {journal}
  {\bibinfo  {journal} {Journal of Statistical Mechanics: Theory and
  Experiment}\ }\textbf {\bibinfo {volume} {2004}},\ \bibinfo {pages} {P04005}
  (\bibinfo {year} {2004})}\BibitemShut {NoStop}
\bibitem [{\citenamefont
  {Vidal}(2004)}]{vidal_tebd_pure_state_dynamics_breakthrough}
  \BibitemOpen
  \bibfield  {author} {\bibinfo {author} {\bibfnamefont {G.}~\bibnamefont
  {Vidal}},\ }\href {\doibase 10.1103/PhysRevLett.93.040502} {\bibfield
  {journal} {\bibinfo  {journal} {Phys. Rev. Lett.}\ }\textbf {\bibinfo
  {volume} {93}},\ \bibinfo {pages} {040502} (\bibinfo {year}
  {2004})}\BibitemShut {NoStop}
\bibitem [{\citenamefont {White}\ and\ \citenamefont
  {Feiguin}(2004)}]{white_feiguin_tdmrg_pure_state_breakthrough}
  \BibitemOpen
  \bibfield  {author} {\bibinfo {author} {\bibfnamefont {S.~R.}\ \bibnamefont
  {White}}\ and\ \bibinfo {author} {\bibfnamefont {A.~E.}\ \bibnamefont
  {Feiguin}},\ }\href {\doibase 10.1103/PhysRevLett.93.076401} {\bibfield
  {journal} {\bibinfo  {journal} {Phys. Rev. Lett.}\ }\textbf {\bibinfo
  {volume} {93}},\ \bibinfo {pages} {076401} (\bibinfo {year}
  {2004})}\BibitemShut {NoStop}
\bibitem [{\citenamefont {Schollw{\"o}ck}(2011)}]{schollwock_dmrg_long_notes}
  \BibitemOpen
  \bibfield  {author} {\bibinfo {author} {\bibfnamefont {U.}~\bibnamefont
  {Schollw{\"o}ck}},\ }\href {\doibase 10.1016/j.aop.2010.09.012} {\bibfield
  {journal} {\bibinfo  {journal} {Annals of Physics}\ }\textbf {\bibinfo
  {volume} {326}},\ \bibinfo {pages} {96 } (\bibinfo {year} {2011})},\ \bibinfo
  {note} {january 2011 Special Issue}\BibitemShut {NoStop}
\bibitem [{\citenamefont {Bouillot}\ \emph {et~al.}(2011)\citenamefont
  {Bouillot}, \citenamefont {Kollath}, \citenamefont {L{\"a}uchli},
  \citenamefont {Zvonarev}, \citenamefont {Thielemann}, \citenamefont
  {R{\"u}egg}, \citenamefont {Orignac}, \citenamefont {Citro}, \citenamefont
  {Klanj{\v{s}}ek}, \citenamefont {Berthier}, \citenamefont {Horvati{\'c}},\
  and\ \citenamefont {Giamarchi}}]{bouillot_long_ladder}
  \BibitemOpen
  \bibfield  {author} {\bibinfo {author} {\bibfnamefont {P.}~\bibnamefont
  {Bouillot}}, \bibinfo {author} {\bibfnamefont {C.}~\bibnamefont {Kollath}},
  \bibinfo {author} {\bibfnamefont {A.~M.}\ \bibnamefont {L{\"a}uchli}},
  \bibinfo {author} {\bibfnamefont {M.}~\bibnamefont {Zvonarev}}, \bibinfo
  {author} {\bibfnamefont {B.}~\bibnamefont {Thielemann}}, \bibinfo {author}
  {\bibfnamefont {C.}~\bibnamefont {R{\"u}egg}}, \bibinfo {author}
  {\bibfnamefont {E.}~\bibnamefont {Orignac}}, \bibinfo {author} {\bibfnamefont
  {R.}~\bibnamefont {Citro}}, \bibinfo {author} {\bibfnamefont
  {M.}~\bibnamefont {Klanj{\v{s}}ek}}, \bibinfo {author} {\bibfnamefont
  {C.}~\bibnamefont {Berthier}}, \bibinfo {author} {\bibfnamefont
  {M.}~\bibnamefont {Horvati{\'c}}}, \ and\ \bibinfo {author} {\bibfnamefont
  {T.}~\bibnamefont {Giamarchi}},\ }\href {\doibase 10.1103/PhysRevB.83.054407}
  {\bibfield  {journal} {\bibinfo  {journal} {Physical Review B}\ }\textbf
  {\bibinfo {volume} {83}},\ \bibinfo {pages} {054407} (\bibinfo {year}
  {2011})}\BibitemShut {NoStop}
\bibitem [{\citenamefont {Schmidiger}\ \emph
  {et~al.}(2013{\natexlab{a}})\citenamefont {Schmidiger}, \citenamefont
  {M{\"u}hlbauer}, \citenamefont {Zheludev}, \citenamefont {Bouillot},
  \citenamefont {Giamarchi}, \citenamefont {Kollath}, \citenamefont {Ehlers},\
  and\ \citenamefont {Tsvelik}}]{schmidiger_neutrons_bound_spinons}
  \BibitemOpen
  \bibfield  {author} {\bibinfo {author} {\bibfnamefont {D.}~\bibnamefont
  {Schmidiger}}, \bibinfo {author} {\bibfnamefont {S.}~\bibnamefont
  {M{\"u}hlbauer}}, \bibinfo {author} {\bibfnamefont {A.}~\bibnamefont
  {Zheludev}}, \bibinfo {author} {\bibfnamefont {P.}~\bibnamefont {Bouillot}},
  \bibinfo {author} {\bibfnamefont {T.}~\bibnamefont {Giamarchi}}, \bibinfo
  {author} {\bibfnamefont {C.}~\bibnamefont {Kollath}}, \bibinfo {author}
  {\bibfnamefont {G.}~\bibnamefont {Ehlers}}, \ and\ \bibinfo {author}
  {\bibfnamefont {A.~M.}\ \bibnamefont {Tsvelik}},\ }\href 
{\doibase  10.1103/PhysRevB.88.094411} {\bibfield  {journal} {\bibinfo  {journal}
  {Physical Review B}\ }\textbf {\bibinfo {volume} {88}},\ \bibinfo {pages}
  {094411} (\bibinfo {year} {2013}{\natexlab{a}})}\BibitemShut {NoStop}
\bibitem [{\citenamefont {Klanj{\v{s}}ek}\ \emph {et~al.}(2008)\citenamefont
  {Klanj{\v{s}}ek}, \citenamefont {Mayaffre}, \citenamefont {Berthier},
  \citenamefont {Horvati{\'c}}, \citenamefont {Chiari}, \citenamefont
  {Piovesana}, \citenamefont {Bouillot}, \citenamefont {Kollath}, \citenamefont
  {Orignac}, \citenamefont {Citro},\ and\ \citenamefont
  {Giamarchi}}]{klanjsek_bpcp}
  \BibitemOpen
  \bibfield  {author} {\bibinfo {author} {\bibfnamefont {M.}~\bibnamefont
  {Klanj{\v{s}}ek}}, \bibinfo {author} {\bibfnamefont {H.}~\bibnamefont
  {Mayaffre}}, \bibinfo {author} {\bibfnamefont {C.}~\bibnamefont {Berthier}},
  \bibinfo {author} {\bibfnamefont {M.}~\bibnamefont {Horvati{\'c}}}, \bibinfo
  {author} {\bibfnamefont {B.}~\bibnamefont {Chiari}}, \bibinfo {author}
  {\bibfnamefont {O.}~\bibnamefont {Piovesana}}, \bibinfo {author}
  {\bibfnamefont {P.}~\bibnamefont {Bouillot}}, \bibinfo {author}
  {\bibfnamefont {C.}~\bibnamefont {Kollath}}, \bibinfo {author} {\bibfnamefont
  {E.}~\bibnamefont {Orignac}}, \bibinfo {author} {\bibfnamefont
  {R.}~\bibnamefont {Citro}}, \ and\ \bibinfo {author} {\bibfnamefont
  {T.}~\bibnamefont {Giamarchi}},\ }\href 
{\doibase  10.1103/PhysRevLett.101.137207} {\bibfield  {journal} {\bibinfo  {journal}
  {Physical Review Letters}\ }\textbf {\bibinfo {volume} {101}},\ \bibinfo
  {pages} {137207} (\bibinfo {year} {2008})}\BibitemShut {NoStop}
\bibitem [{\citenamefont {Schmidiger}\ \emph {et~al.}(2012)\citenamefont
  {Schmidiger}, \citenamefont {Bouillot}, \citenamefont {M{\"u}hlbauer},
  \citenamefont {Gvasaliya}, \citenamefont {Kollath}, \citenamefont
  {Giamarchi},\ and\ \citenamefont {Zheludev}}]{schmidiger_dimpy_neutrons}
  \BibitemOpen
  \bibfield  {author} {\bibinfo {author} {\bibfnamefont {D.}~\bibnamefont
  {Schmidiger}}, \bibinfo {author} {\bibfnamefont {P.}~\bibnamefont
  {Bouillot}}, \bibinfo {author} {\bibfnamefont {S.}~\bibnamefont
  {M{\"u}hlbauer}}, \bibinfo {author} {\bibfnamefont {S.}~\bibnamefont
  {Gvasaliya}}, \bibinfo {author} {\bibfnamefont {C.}~\bibnamefont {Kollath}},
  \bibinfo {author} {\bibfnamefont {T.}~\bibnamefont {Giamarchi}}, \ and\
  \bibinfo {author} {\bibfnamefont {A.}~\bibnamefont {Zheludev}},\ }\href
  {\doibase 10.1103/PhysRevLett.108.167201} {\bibfield  {journal} {\bibinfo
  {journal} {Physical Review Letters}\ }\textbf {\bibinfo {volume} {108}},\
  \bibinfo {pages} {167201} (\bibinfo {year} {2012})}\BibitemShut {NoStop}
\bibitem [{\citenamefont {Sachdev}(1999)}]{sachdev_qcp_book}
  \BibitemOpen
  \bibfield  {author} {\bibinfo {author} {\bibfnamefont {S.}~\bibnamefont
  {Sachdev}},\ }\href@noop {} {\emph {\bibinfo {title} {Quantum Phase
  Transitions}}}\ (\bibinfo  {publisher} {Cambridge University Press},\
  \bibinfo {year} {1999})\BibitemShut {NoStop}
\bibitem [{\citenamefont {Sachdev}\ \emph {et~al.}(1994)\citenamefont
  {Sachdev}, \citenamefont {Senthil},\ and\ \citenamefont
  {Shankar}}]{sachdev_qcp_antiferro_finiteT}
  \BibitemOpen
  \bibfield  {author} {\bibinfo {author} {\bibfnamefont {S.}~\bibnamefont
  {Sachdev}}, \bibinfo {author} {\bibfnamefont {T.}~\bibnamefont {Senthil}}, \
  and\ \bibinfo {author} {\bibfnamefont {R.}~\bibnamefont {Shankar}},\ }\href
  {\doibase 10.1103/PhysRevB.50.258} {\bibfield  {journal} {\bibinfo  {journal}
  {Physical Review B}\ }\textbf {\bibinfo {volume} {50}},\ \bibinfo {pages}
  {258} (\bibinfo {year} {1994})}\BibitemShut {NoStop}
\bibitem [{\citenamefont {Verstraete}\ \emph {et~al.}(2004)\citenamefont
  {Verstraete}, \citenamefont {Garc{\'i}a-Ripoll},\ and\ \citenamefont
  {Cirac}}]{verstraete_garciaripoll_cirac_tdmrg_dissipative_breakthrough}
  \BibitemOpen
  \bibfield  {author} {\bibinfo {author} {\bibfnamefont {F.}~\bibnamefont
  {Verstraete}}, \bibinfo {author} {\bibfnamefont {J.~J.}\ \bibnamefont
  {Garc{\'i}a-Ripoll}}, \ and\ \bibinfo {author} {\bibfnamefont {J.~I.}\
  \bibnamefont {Cirac}},\ }\href {\doibase 10.1103/PhysRevLett.93.207204}
  {\bibfield  {journal} {\bibinfo  {journal} {Phys. Rev. Lett.}\ }\textbf
  {\bibinfo {volume} {93}},\ \bibinfo {pages} {207204} (\bibinfo {year}
  {2004})}\BibitemShut {NoStop}
\bibitem [{\citenamefont {Zwolak}\ and\ \citenamefont
  {Vidal}(2004)}]{vidal_zwolak_tebd_temperature}
  \BibitemOpen
  \bibfield  {author} {\bibinfo {author} {\bibfnamefont {M.}~\bibnamefont
  {Zwolak}}\ and\ \bibinfo {author} {\bibfnamefont {G.}~\bibnamefont {Vidal}},\
  }\href {\doibase 10.1103/PhysRevLett.93.207205} {\bibfield  {journal}
  {\bibinfo  {journal} {Phys. Rev. Lett.}\ }\textbf {\bibinfo {volume} {93}},\
  \bibinfo {pages} {207205} (\bibinfo {year} {2004})}\BibitemShut {NoStop}
\bibitem [{\citenamefont {Feiguin}\ and\ \citenamefont
  {White}(2005)}]{feiguin_white_Tdmrg_just_after}
  \BibitemOpen
  \bibfield  {author} {\bibinfo {author} {\bibfnamefont {A.~E.}\ \bibnamefont
  {Feiguin}}\ and\ \bibinfo {author} {\bibfnamefont {S.~R.}\ \bibnamefont
  {White}},\ }\href {\doibase 10.1103/PhysRevB.72.220401} {\bibfield  {journal}
  {\bibinfo  {journal} {Phys. Rev. B}\ }\textbf {\bibinfo {volume} {72}},\
  \bibinfo {pages} {220401} (\bibinfo {year} {2005})}\BibitemShut {NoStop}
\bibitem [{\citenamefont {Barthel}(2013)}]{barthel_tdmrg_optimized_schemes}
  \BibitemOpen
  \bibfield  {author} {\bibinfo {author} {\bibfnamefont {T.}~\bibnamefont
  {Barthel}},\ }\href {\doibase 10.1088/1367-2630/15/7/073010} {\bibfield
  {journal} {\bibinfo  {journal} {New Journal of Physics}\ }\textbf {\bibinfo
  {volume} {15}},\ \bibinfo {pages} {073010} (\bibinfo {year}
  {2013})}\BibitemShut {NoStop}
\bibitem [{\citenamefont {Barthel}\ \emph {et~al.}(2009)\citenamefont
  {Barthel}, \citenamefont {Schollw{\"o}ck},\ and\ \citenamefont
  {White}}]{barthel_xxz_Tdmrg}
  \BibitemOpen
  \bibfield  {author} {\bibinfo {author} {\bibfnamefont {T.}~\bibnamefont
  {Barthel}}, \bibinfo {author} {\bibfnamefont {U.}~\bibnamefont
  {Schollw{\"o}ck}}, \ and\ \bibinfo {author} {\bibfnamefont {S.~R.}\
  \bibnamefont {White}},\ }\href {\doibase 10.1103/PhysRevB.79.245101}
  {\bibfield  {journal} {\bibinfo  {journal} {Physical Review B}\ }\textbf
  {\bibinfo {volume} {79}},\ \bibinfo {pages} {245101} (\bibinfo {year}
  {2009})}\BibitemShut {NoStop}
\bibitem [{\citenamefont {Blosser}\ \emph {et~al.}(2017)\citenamefont
  {Blosser}, \citenamefont {Kestin}, \citenamefont {Povarov}, \citenamefont
  {Bewley}, \citenamefont {Coira}, \citenamefont {Giamarchi},\ and\
  \citenamefont {Zheludev}}]{noam_quasi1D}
  \BibitemOpen
  \bibfield  {author} {\bibinfo {author} {\bibfnamefont {D.}~\bibnamefont
  {Blosser}}, \bibinfo {author} {\bibfnamefont {N.}~\bibnamefont {Kestin}},
  \bibinfo {author} {\bibfnamefont {K.~Y.}\ \bibnamefont {Povarov}}, \bibinfo
  {author} {\bibfnamefont {R.}~\bibnamefont {Bewley}}, \bibinfo {author}
  {\bibfnamefont {E.}~\bibnamefont {Coira}}, \bibinfo {author} {\bibfnamefont
  {T.}~\bibnamefont {Giamarchi}}, \ and\ \bibinfo {author} {\bibfnamefont
  {A.}~\bibnamefont {Zheludev}},\ }\href {\doibase 10.1103/PhysRevB.96.134406}
  {\bibfield  {journal} {\bibinfo  {journal} {Physical Review B}\ }\textbf
  {\bibinfo {volume} {96}},\ \bibinfo {pages} {134406} (\bibinfo {year}
  {2017})}\BibitemShut {NoStop}
\bibitem [{\citenamefont {Becker}\ \emph {et~al.}(2017)\citenamefont {Becker},
  \citenamefont {K{\"o}hler}, \citenamefont {Tiegel}, \citenamefont {Manmana},
  \citenamefont {Wessel},\ and\ \citenamefont
  {Honecker}}]{honecker_spinone_Tdmrg}
  \BibitemOpen
  \bibfield  {author} {\bibinfo {author} {\bibfnamefont {J.}~\bibnamefont
  {Becker}}, \bibinfo {author} {\bibfnamefont {T.}~\bibnamefont {K{\"o}hler}},
  \bibinfo {author} {\bibfnamefont {A.~C.}\ \bibnamefont {Tiegel}}, \bibinfo
  {author} {\bibfnamefont {S.~R.}\ \bibnamefont {Manmana}}, \bibinfo {author}
  {\bibfnamefont {S.}~\bibnamefont {Wessel}}, \ and\ \bibinfo {author}
  {\bibfnamefont {A.}~\bibnamefont {Honecker}},\ }\href 
{\doibase  10.1103/PhysRevB.96.060403} {\bibfield  {journal} {\bibinfo  {journal}
  {Physical Review B}\ }\textbf {\bibinfo {volume} {96}},\ \bibinfo {pages}
  {060403} (\bibinfo {year} {2017})}\BibitemShut {NoStop}
\bibitem [{\citenamefont {Lange}\ \emph {et~al.}(2018)\citenamefont {Lange},
  \citenamefont {Ejima},\ and\ \citenamefont {Fehske}}]{lange_spinone_tdmrg}
  \BibitemOpen
  \bibfield  {author} {\bibinfo {author} {\bibfnamefont {F.}~\bibnamefont
  {Lange}}, \bibinfo {author} {\bibfnamefont {S.}~\bibnamefont {Ejima}}, \ and\
  \bibinfo {author} {\bibfnamefont {H.}~\bibnamefont {Fehske}},\ }\href
  {\doibase 10.1103/PhysRevB.97.060403} {\bibfield  {journal} {\bibinfo
  {journal} {Physical Review B}\ }\textbf {\bibinfo {volume} {97}},\ \bibinfo
  {pages} {060403} (\bibinfo {year} {2018})}\BibitemShut {NoStop}
\bibitem [{\citenamefont {Klyushina}\ \emph {et~al.}(2016)\citenamefont
  {Klyushina}, \citenamefont {Tiegel}, \citenamefont {Fauseweh}, \citenamefont
  {Islam}, \citenamefont {Park}, \citenamefont {Klemke}, \citenamefont
  {Honecker}, \citenamefont {Uhrig}, \citenamefont {Manmana},\ and\
  \citenamefont {Lake}}]{klyushina_dimer_Tdmrg}
  \BibitemOpen
  \bibfield  {author} {\bibinfo {author} {\bibfnamefont {E.~S.}\ \bibnamefont
  {Klyushina}}, \bibinfo {author} {\bibfnamefont {A.~C.}\ \bibnamefont
  {Tiegel}}, \bibinfo {author} {\bibfnamefont {B.}~\bibnamefont {Fauseweh}},
  \bibinfo {author} {\bibfnamefont {A.~T. M.~N.}\ \bibnamefont {Islam}},
  \bibinfo {author} {\bibfnamefont {J.~T.}\ \bibnamefont {Park}}, \bibinfo
  {author} {\bibfnamefont {B.}~\bibnamefont {Klemke}}, \bibinfo {author}
  {\bibfnamefont {A.}~\bibnamefont {Honecker}}, \bibinfo {author}
  {\bibfnamefont {G.~S.}\ \bibnamefont {Uhrig}}, \bibinfo {author}
  {\bibfnamefont {S.~R.}\ \bibnamefont {Manmana}}, \ and\ \bibinfo {author}
  {\bibfnamefont {B.}~\bibnamefont {Lake}},\ }\href 
{\doibase  10.1103/PhysRevB.93.241109} {\bibfield  {journal} {\bibinfo  {journal}
  {Physical Review B}\ }\textbf {\bibinfo {volume} {93}},\ \bibinfo {pages}
  {241109} (\bibinfo {year} {2016})}\BibitemShut {NoStop}
\bibitem [{\citenamefont {Coira}\ \emph {et~al.}(2016)\citenamefont {Coira},
  \citenamefont {Barmettler}, \citenamefont {Giamarchi},\ and\ \citenamefont
  {Kollath}}]{coira_giam_kollath_spin_lattice_relax_nmr_low_t}
  \BibitemOpen
  \bibfield  {author} {\bibinfo {author} {\bibfnamefont {E.}~\bibnamefont
  {Coira}}, \bibinfo {author} {\bibfnamefont {P.}~\bibnamefont {Barmettler}},
  \bibinfo {author} {\bibfnamefont {T.}~\bibnamefont {Giamarchi}}, \ and\
  \bibinfo {author} {\bibfnamefont {C.}~\bibnamefont {Kollath}},\ }\href
  {\doibase 10.1103/PhysRevB.94.144408} {\bibfield  {journal} {\bibinfo
  {journal} {Physical Review B}\ }\textbf {\bibinfo {volume} {94}},\ \bibinfo
  {pages} {144408} (\bibinfo {year} {2016})}\BibitemShut {NoStop}
\bibitem [{\citenamefont {Coira}\ \emph {et~al.}(2018)\citenamefont {Coira},
  \citenamefont {Barmettler}, \citenamefont {Giamarchi},\ and\ \citenamefont
  {Kollath}}]{coira_temperature_dimer}
  \BibitemOpen
  \bibfield  {author} {\bibinfo {author} {\bibfnamefont {E.}~\bibnamefont
  {Coira}}, \bibinfo {author} {\bibfnamefont {P.}~\bibnamefont {Barmettler}},
  \bibinfo {author} {\bibfnamefont {T.}~\bibnamefont {Giamarchi}}, \ and\
  \bibinfo {author} {\bibfnamefont {C.}~\bibnamefont {Kollath}},\ }\href
  {\doibase 10.1103/PhysRevB.98.104435} {\bibfield  {journal} {\bibinfo
  {journal} {Physical Review B}\ }\textbf {\bibinfo {volume} {98}},\ \bibinfo
  {pages} {104435} (\bibinfo {year} {2018})}\BibitemShut {NoStop}
\bibitem [{\citenamefont {Damle}\ and\ \citenamefont
  {Sachdev}(1998)}]{damle_temperature_spin_dynamics}
  \BibitemOpen
  \bibfield  {author} {\bibinfo {author} {\bibfnamefont {K.}~\bibnamefont
  {Damle}}\ and\ \bibinfo {author} {\bibfnamefont {S.}~\bibnamefont
  {Sachdev}},\ }\href {\doibase 10.1103/PhysRevB.57.8307} {\bibfield  {journal}
  {\bibinfo  {journal} {Physical Review B}\ }\textbf {\bibinfo {volume} {57}},\
  \bibinfo {pages} {8307} (\bibinfo {year} {1998})}\BibitemShut {NoStop}
\bibitem [{\citenamefont {Blosser}\ \emph {et~al.}(2018)\citenamefont
  {Blosser}, \citenamefont {Bhartiya}, \citenamefont {Voneshen},\ and\
  \citenamefont {Zheludev}}]{blosser_qcp_ladder}
  \BibitemOpen
  \bibfield  {author} {\bibinfo {author} {\bibfnamefont {D.}~\bibnamefont
  {Blosser}}, \bibinfo {author} {\bibfnamefont {V.~K.}\ \bibnamefont
  {Bhartiya}}, \bibinfo {author} {\bibfnamefont {D.~J.}\ \bibnamefont
  {Voneshen}}, \ and\ \bibinfo {author} {\bibfnamefont {A.}~\bibnamefont
  {Zheludev}},\ }\href {\doibase 10.1103/PhysRevLett.121.247201} {\bibfield
  {journal} {\bibinfo  {journal} {Physical Review Letters}\ }\textbf {\bibinfo
  {volume} {121}},\ \bibinfo {pages} {247201} (\bibinfo {year}
  {2018})}\BibitemShut {NoStop}
\bibitem [{\citenamefont {Ward}\ \emph {et~al.}(2013)\citenamefont {Ward},
  \citenamefont {Bouillot}, \citenamefont {Ryll}, \citenamefont {Kiefer},
  \citenamefont {Kr{\"a}mer}, \citenamefont {R{\"u}egg}, \citenamefont
  {Kollath},\ and\ \citenamefont {Giamarchi}}]{ward_bpcc_compound}
  \BibitemOpen
  \bibfield  {author} {\bibinfo {author} {\bibfnamefont {S.}~\bibnamefont
  {Ward}}, \bibinfo {author} {\bibfnamefont {P.}~\bibnamefont {Bouillot}},
  \bibinfo {author} {\bibfnamefont {H.}~\bibnamefont {Ryll}}, \bibinfo {author}
  {\bibfnamefont {K.}~\bibnamefont {Kiefer}}, \bibinfo {author} {\bibfnamefont
  {K.~W.}\ \bibnamefont {Kr{\"a}mer}}, \bibinfo {author} {\bibfnamefont
  {C.}~\bibnamefont {R{\"u}egg}}, \bibinfo {author} {\bibfnamefont
  {C.}~\bibnamefont {Kollath}}, \ and\ \bibinfo {author} {\bibfnamefont
  {T.}~\bibnamefont {Giamarchi}},\ }\href 
{\doibase  10.1088/0953-8984/25/1/014004} {\bibfield  {journal} {\bibinfo  {journal}
  {Journal of Physics. Condensed Matter: An Institute of Physics Journal}\
  }\textbf {\bibinfo {volume} {25}},\ \bibinfo {pages} {014004} (\bibinfo
  {year} {2013})}\BibitemShut {NoStop}
\bibitem [{\citenamefont {Ward}\ \emph {et~al.}(2017)\citenamefont {Ward},
  \citenamefont {Mena}, \citenamefont {Bouillot}, \citenamefont {Kollath},
  \citenamefont {Giamarchi}, \citenamefont {Schmidt}, \citenamefont {Normand},
  \citenamefont {Kr{\"a}mer}, \citenamefont {Biner}, \citenamefont {Bewley},
  \citenamefont {Guidi}, \citenamefont {Boehm}, \citenamefont {McMorrow},\ and\
  \citenamefont {R{\"u}egg}}]{ward_bpcc_singlet}
  \BibitemOpen
  \bibfield  {author} {\bibinfo {author} {\bibfnamefont {S.}~\bibnamefont
  {Ward}}, \bibinfo {author} {\bibfnamefont {M.}~\bibnamefont {Mena}}, \bibinfo
  {author} {\bibfnamefont {P.}~\bibnamefont {Bouillot}}, \bibinfo {author}
  {\bibfnamefont {C.}~\bibnamefont {Kollath}}, \bibinfo {author} {\bibfnamefont
  {T.}~\bibnamefont {Giamarchi}}, \bibinfo {author} {\bibfnamefont {K.~P.}\
  \bibnamefont {Schmidt}}, \bibinfo {author} {\bibfnamefont {B.}~\bibnamefont
  {Normand}}, \bibinfo {author} {\bibfnamefont {K.~W.}\ \bibnamefont
  {Kr{\"a}mer}}, \bibinfo {author} {\bibfnamefont {D.}~\bibnamefont {Biner}},
  \bibinfo {author} {\bibfnamefont {R.}~\bibnamefont {Bewley}}, \bibinfo
  {author} {\bibfnamefont {T.}~\bibnamefont {Guidi}}, \bibinfo {author}
  {\bibfnamefont {M.}~\bibnamefont {Boehm}}, \bibinfo {author} {\bibfnamefont
  {D.~F.}\ \bibnamefont {McMorrow}}, \ and\ \bibinfo {author} {\bibfnamefont
  {C.}~\bibnamefont {R{\"u}egg}},\ }\href 
{\doibase  10.1103/PhysRevLett.118.177202} {\bibfield  {journal} {\bibinfo  {journal}
  {Physical Review Letters}\ }\textbf {\bibinfo {volume} {118}},\ \bibinfo
  {pages} {177202} (\bibinfo {year} {2017})}\BibitemShut {NoStop}
\bibitem [{\citenamefont {Ryll}\ \emph {et~al.}(2014)\citenamefont {Ryll},
  \citenamefont {Kiefer}, \citenamefont {R{\"u}egg}, \citenamefont {Ward},
  \citenamefont {Kr{\"a}mer}, \citenamefont {Biner}, \citenamefont {Bouillot},
  \citenamefont {Coira}, \citenamefont {Giamarchi},\ and\ \citenamefont
  {Kollath}}]{ryll_gruneisen_bpcc}
  \BibitemOpen
  \bibfield  {author} {\bibinfo {author} {\bibfnamefont {H.}~\bibnamefont
  {Ryll}}, \bibinfo {author} {\bibfnamefont {K.}~\bibnamefont {Kiefer}},
  \bibinfo {author} {\bibfnamefont {C.}~\bibnamefont {R{\"u}egg}}, \bibinfo
  {author} {\bibfnamefont {S.}~\bibnamefont {Ward}}, \bibinfo {author}
  {\bibfnamefont {K.~W.}\ \bibnamefont {Kr{\"a}mer}}, \bibinfo {author}
  {\bibfnamefont {D.}~\bibnamefont {Biner}}, \bibinfo {author} {\bibfnamefont
  {P.}~\bibnamefont {Bouillot}}, \bibinfo {author} {\bibfnamefont
  {E.}~\bibnamefont {Coira}}, \bibinfo {author} {\bibfnamefont
  {T.}~\bibnamefont {Giamarchi}}, \ and\ \bibinfo {author} {\bibfnamefont
  {C.}~\bibnamefont {Kollath}},\ }\href {\doibase 10.1103/PhysRevB.89.144416}
  {\bibfield  {journal} {\bibinfo  {journal} {Physical Review B}\ }\textbf
  {\bibinfo {volume} {89}},\ \bibinfo {pages} {144416} (\bibinfo {year}
  {2014})}\BibitemShut {NoStop}
\bibitem [{\citenamefont {Tajiri}\ \emph {et~al.}(2004)\citenamefont {Tajiri},
  \citenamefont {Deguchi}, \citenamefont {Mito}, \citenamefont {Takagi},
  \citenamefont {Nojiri}, \citenamefont {Kawae},\ and\ \citenamefont
  {Takeda}}]{tajiri_magnetisation_bpcc}
  \BibitemOpen
  \bibfield  {author} {\bibinfo {author} {\bibfnamefont {T.}~\bibnamefont
  {Tajiri}}, \bibinfo {author} {\bibfnamefont {H.}~\bibnamefont {Deguchi}},
  \bibinfo {author} {\bibfnamefont {M.}~\bibnamefont {Mito}}, \bibinfo {author}
  {\bibfnamefont {S.}~\bibnamefont {Takagi}}, \bibinfo {author} {\bibfnamefont
  {H.}~\bibnamefont {Nojiri}}, \bibinfo {author} {\bibfnamefont
  {T.}~\bibnamefont {Kawae}}, \ and\ \bibinfo {author} {\bibfnamefont
  {K.}~\bibnamefont {Takeda}},\ }\href {\doibase 10.1016/j.jmmm.2003.12.014}
  {\bibfield  {journal} {\bibinfo  {journal} {Journal of Magnetism and Magnetic
  Materials}\ }\bibinfo {series} {Proceedings of the {International}
  {Conference} on {Magnetism} ({ICM} 2003)},\ \textbf {\bibinfo {volume}
  {272-276}},\ \bibinfo {pages} {1070} (\bibinfo {year} {2004})}\BibitemShut
  {NoStop}
\bibitem [{\citenamefont {Suzuki}(1991)}]{suzuki_fractalization_exp_gates}
  \BibitemOpen
  \bibfield  {author} {\bibinfo {author} {\bibfnamefont {M.}~\bibnamefont
  {Suzuki}},\ }\href {\doibase 10.1063/1.529425} {\bibfield  {journal}
  {\bibinfo  {journal} {Journal of Mathematical Physics}\ }\textbf {\bibinfo
  {volume} {32}},\ \bibinfo {pages} {400} (\bibinfo {year} {1991})}\BibitemShut
  {NoStop}
\bibitem [{\citenamefont {McLachlan}(1995)}]{mclachlan_leapfrog_exp}
  \BibitemOpen
  \bibfield  {author} {\bibinfo {author} {\bibfnamefont {R.}~\bibnamefont
  {McLachlan}},\ }\href {\doibase 10.1137/0916010} {\bibfield  {journal}
  {\bibinfo  {journal} {SIAM Journal on Scientific Computing}\ }\textbf
  {\bibinfo {volume} {16}},\ \bibinfo {pages} {151} (\bibinfo {year}
  {1995})}\BibitemShut {NoStop}
\bibitem [{\citenamefont {Singh}\ \emph {et~al.}(2011)\citenamefont {Singh},
  \citenamefont {Pfeifer},\ and\ \citenamefont
  {Vidal}}]{singh_qn_dmrg_tensornetwork}
  \BibitemOpen
  \bibfield  {author} {\bibinfo {author} {\bibfnamefont {S.}~\bibnamefont
  {Singh}}, \bibinfo {author} {\bibfnamefont {R.~N.~C.}\ \bibnamefont
  {Pfeifer}}, \ and\ \bibinfo {author} {\bibfnamefont {G.}~\bibnamefont
  {Vidal}},\ }\href {\doibase 10.1103/PhysRevB.83.115125} {\bibfield  {journal}
  {\bibinfo  {journal} {Physical Review B}\ }\textbf {\bibinfo {volume} {83}},\
  \bibinfo {pages} {115125} (\bibinfo {year} {2011})}\BibitemShut {NoStop}
\bibitem [{\citenamefont {Hubig}\ \emph {et~al.}(2017)\citenamefont {Hubig},
  \citenamefont {McCulloch},\ and\ \citenamefont
  {Schollw{\"o}ck}}]{hubig_qn_simple}
  \BibitemOpen
  \bibfield  {author} {\bibinfo {author} {\bibfnamefont {C.}~\bibnamefont
  {Hubig}}, \bibinfo {author} {\bibfnamefont {I.~P.}\ \bibnamefont
  {McCulloch}}, \ and\ \bibinfo {author} {\bibfnamefont {U.}~\bibnamefont
  {Schollw{\"o}ck}},\ }\href {\doibase 10.1103/PhysRevB.95.035129} {\bibfield
  {journal} {\bibinfo  {journal} {Physical Review B}\ }\textbf {\bibinfo
  {volume} {95}},\ \bibinfo {pages} {035129} (\bibinfo {year}
  {2017})}\BibitemShut {NoStop}
\bibitem [{\citenamefont {Eisert}\ \emph {et~al.}(2010)\citenamefont {Eisert},
  \citenamefont {Cramer},\ and\ \citenamefont
  {Plenio}}]{eisert_entanglement_arealaw}
  \BibitemOpen
  \bibfield  {author} {\bibinfo {author} {\bibfnamefont {J.}~\bibnamefont
  {Eisert}}, \bibinfo {author} {\bibfnamefont {M.}~\bibnamefont {Cramer}}, \
  and\ \bibinfo {author} {\bibfnamefont {M.~B.}\ \bibnamefont {Plenio}},\
  }\href {\doibase 10.1103/RevModPhys.82.277} {\bibfield  {journal} {\bibinfo
  {journal} {Reviews of Modern Physics}\ }\textbf {\bibinfo {volume} {82}},\
  \bibinfo {pages} {277} (\bibinfo {year} {2010})}\BibitemShut {NoStop}
\bibitem [{\citenamefont {Tange}(2015)}]{tange_gnuparallel}
  \BibitemOpen
  \bibfield  {author} {\bibinfo {author} {\bibfnamefont {O.}~\bibnamefont
  {Tange}},\ }\href {\doibase 10.5281/zenodo.16303} {\enquote {\bibinfo {title}
  {Gnu parallel 20150322 ('hellwig')},}\ } (\bibinfo {year} {2015}),\ \bibinfo
  {note} {{GNU Parallel is a general parallelizer to run multiple serial
  command line programs in parallel without changing them.}}\BibitemShut
  {Stop}
\bibitem [{\citenamefont {Schulz}\ and\ \citenamefont
  {Bourbonnais}(1983)}]{schulz_quasi1D_sinegordon}
  \BibitemOpen
  \bibfield  {author} {\bibinfo {author} {\bibfnamefont {H.~J.}\ \bibnamefont
  {Schulz}}\ and\ \bibinfo {author} {\bibfnamefont {C.}~\bibnamefont
  {Bourbonnais}},\ }\href {\doibase 10.1103/PhysRevB.27.5856} {\bibfield
  {journal} {\bibinfo  {journal} {Physical Review B}\ }\textbf {\bibinfo
  {volume} {27}},\ \bibinfo {pages} {5856} (\bibinfo {year}
  {1983})}\BibitemShut {NoStop}
\bibitem [{\citenamefont {Chitra}\ and\ \citenamefont
  {Giamarchi}(1997)}]{chitra_giam_qcp_universal_function_ll}
  \BibitemOpen
  \bibfield  {author} {\bibinfo {author} {\bibfnamefont {R.}~\bibnamefont
  {Chitra}}\ and\ \bibinfo {author} {\bibfnamefont {T.}~\bibnamefont
  {Giamarchi}},\ }\href {\doibase 10.1103/PhysRevB.55.5816} {\bibfield
  {journal} {\bibinfo  {journal} {Physical Review B}\ }\textbf {\bibinfo
  {volume} {55}},\ \bibinfo {pages} {5816} (\bibinfo {year}
  {1997})}\BibitemShut {NoStop}
\bibitem [{\citenamefont {Giamarchi}\ and\ \citenamefont
  {Tsvelik}(1999)}]{giam_tsvelik_qcp_ladders_interplay_coupling}
  \BibitemOpen
  \bibfield  {author} {\bibinfo {author} {\bibfnamefont {T.}~\bibnamefont
  {Giamarchi}}\ and\ \bibinfo {author} {\bibfnamefont {A.~M.}\ \bibnamefont
  {Tsvelik}},\ }\href {\doibase 10.1103/PhysRevB.59.11398} {\bibfield
  {journal} {\bibinfo  {journal} {Physical Review B}\ }\textbf {\bibinfo
  {volume} {59}},\ \bibinfo {pages} {11398} (\bibinfo {year}
  {1999})}\BibitemShut {NoStop}
\bibitem [{\citenamefont {Hikihara}\ and\ \citenamefont
  {Furusaki}(2001)}]{hikihara_furusaki}
  \BibitemOpen
  \bibfield  {author} {\bibinfo {author} {\bibfnamefont {T.}~\bibnamefont
  {Hikihara}}\ and\ \bibinfo {author} {\bibfnamefont {A.}~\bibnamefont
  {Furusaki}},\ }\href {\doibase 10.1103/PhysRevB.63.134438} {\bibfield
  {journal} {\bibinfo  {journal} {Physical Review B}\ }\textbf {\bibinfo
  {volume} {63}},\ \bibinfo {pages} {134438} (\bibinfo {year}
  {2001})}\BibitemShut {NoStop}
\bibitem [{\citenamefont {Dolfi}\ \emph {et~al.}(2014)\citenamefont {Dolfi},
  \citenamefont {Bauer}, \citenamefont {Keller}, \citenamefont {Kosenkov},
  \citenamefont {Ewart}, \citenamefont {Kantian}, \citenamefont {Giamarchi},\
  and\ \citenamefont {Troyer}}]{alps_mps}
  \BibitemOpen
  \bibfield  {author} {\bibinfo {author} {\bibfnamefont {M.}~\bibnamefont
  {Dolfi}}, \bibinfo {author} {\bibfnamefont {B.}~\bibnamefont {Bauer}},
  \bibinfo {author} {\bibfnamefont {S.}~\bibnamefont {Keller}}, \bibinfo
  {author} {\bibfnamefont {A.}~\bibnamefont {Kosenkov}}, \bibinfo {author}
  {\bibfnamefont {T.}~\bibnamefont {Ewart}}, \bibinfo {author} {\bibfnamefont
  {A.}~\bibnamefont {Kantian}}, \bibinfo {author} {\bibfnamefont
  {T.}~\bibnamefont {Giamarchi}}, \ and\ \bibinfo {author} {\bibfnamefont
  {M.}~\bibnamefont {Troyer}},\ }\href {\doibase 10.1016/j.cpc.2014.08.019}
  {\bibfield  {journal} {\bibinfo  {journal} {Computer Physics Communications}\
  }\textbf {\bibinfo {volume} {185}},\ \bibinfo {pages} {3430} (\bibinfo {year}
  {2014})}\BibitemShut {NoStop}
\bibitem [{\citenamefont {R{\"u}egg}\ \emph {et~al.}(2008)\citenamefont
  {R{\"u}egg}, \citenamefont {Kiefer}, \citenamefont {Thielemann},
  \citenamefont {McMorrow}, \citenamefont {Zapf}, \citenamefont {Normand},
  \citenamefont {Zvonarev}, \citenamefont {Bouillot}, \citenamefont {Kollath},
  \citenamefont {Giamarchi}, \citenamefont {Capponi}, \citenamefont
  {Poilblanc}, \citenamefont {Biner},\ and\ \citenamefont
  {Kr{\"a}mer}}]{thielemann_bpcb_neutrons_luttinger}
  \BibitemOpen
  \bibfield  {author} {\bibinfo {author} {\bibfnamefont {C.}~\bibnamefont
  {R{\"u}egg}}, \bibinfo {author} {\bibfnamefont {K.}~\bibnamefont {Kiefer}},
  \bibinfo {author} {\bibfnamefont {B.}~\bibnamefont {Thielemann}}, \bibinfo
  {author} {\bibfnamefont {D.~F.}\ \bibnamefont {McMorrow}}, \bibinfo {author}
  {\bibfnamefont {V.}~\bibnamefont {Zapf}}, \bibinfo {author} {\bibfnamefont
  {B.}~\bibnamefont {Normand}}, \bibinfo {author} {\bibfnamefont {M.~B.}\
  \bibnamefont {Zvonarev}}, \bibinfo {author} {\bibfnamefont {P.}~\bibnamefont
  {Bouillot}}, \bibinfo {author} {\bibfnamefont {C.}~\bibnamefont {Kollath}},
  \bibinfo {author} {\bibfnamefont {T.}~\bibnamefont {Giamarchi}}, \bibinfo
  {author} {\bibfnamefont {S.}~\bibnamefont {Capponi}}, \bibinfo {author}
  {\bibfnamefont {D.}~\bibnamefont {Poilblanc}}, \bibinfo {author}
  {\bibfnamefont {D.}~\bibnamefont {Biner}}, \ and\ \bibinfo {author}
  {\bibfnamefont {K.~W.}\ \bibnamefont {Kr{\"a}mer}},\ }\href 
{\doibase  10.1103/PhysRevLett.101.247202} {\bibfield  {journal} {\bibinfo  {journal}
  {Physical Review Letters}\ }\textbf {\bibinfo {volume} {101}},\ \bibinfo
  {pages} {247202} (\bibinfo {year} {2008})}\BibitemShut {NoStop}
\bibitem [{\citenamefont {Schmidiger}\ \emph
  {et~al.}(2013{\natexlab{b}})\citenamefont {Schmidiger}, \citenamefont
  {Bouillot}, \citenamefont {Guidi}, \citenamefont {Bewley}, \citenamefont
  {Kollath}, \citenamefont {Giamarchi},\ and\ \citenamefont
  {Zheludev}}]{schmidiger_prl2013}
  \BibitemOpen
  \bibfield  {author} {\bibinfo {author} {\bibfnamefont {D.}~\bibnamefont
  {Schmidiger}}, \bibinfo {author} {\bibfnamefont {P.}~\bibnamefont
  {Bouillot}}, \bibinfo {author} {\bibfnamefont {T.}~\bibnamefont {Guidi}},
  \bibinfo {author} {\bibfnamefont {R.}~\bibnamefont {Bewley}}, \bibinfo
  {author} {\bibfnamefont {C.}~\bibnamefont {Kollath}}, \bibinfo {author}
  {\bibfnamefont {T.}~\bibnamefont {Giamarchi}}, \ and\ \bibinfo {author}
  {\bibfnamefont {A.}~\bibnamefont {Zheludev}},\ }\href 
{\doibase  10.1103/PhysRevLett.111.107202} {\bibfield  {journal} {\bibinfo  {journal}
  {Physical Review Letters}\ }\textbf {\bibinfo {volume} {111}},\ \bibinfo
  {pages} {107202} (\bibinfo {year} {2013}{\natexlab{b}})}\BibitemShut
  {NoStop}
\end{thebibliography}
\end{document}